\documentclass[letterpaper,12pt]{article}
\usepackage[margin=1in]{geometry}

\usepackage{amsthm,amsmath,amsfonts,amssymb}
\usepackage[normalem]{ulem}
\usepackage{enumerate}
\usepackage[title,toc,titletoc]{appendix}
\usepackage{graphicx}
\usepackage{setspace}
\usepackage{subcaption}
\usepackage{multirow}
\usepackage{floatrow}
\usepackage{hyperref}
\usepackage{bm}
\usepackage{bbm}
\usepackage{wrapfig}
\usepackage[linesnumbered,ruled,vlined]{algorithm2e}
\usepackage{caption}

\newfloatcommand{capbtabbox}{table}[][\FBwidth]

\usepackage{natbib}
\usepackage{blindtext}

\usepackage{graphicx}
\usepackage{float}
\graphicspath{{./figures/}}

\def\spacingset#1{\renewcommand{\baselinestretch}%
	{#1}\small\normalsize} \spacingset{1}

\usepackage[dvipsnames]{xcolor}

\usepackage{url}

\usepackage{bm} % For bold math symbols
\newcommand{\phib}{\bm{\phi}}

\newcommand{\betab}{\bm{\beta}}

\newcommand{\thetab}{\bm{\theta}}  %% theta vector
\newcommand{\Thetab}{\bm{\Theta}} %%vector theta domain

\newcommand{\bb}{\bm{b}}

\newcommand{\db}{\bm{d}}

\newcommand{\hb}{\bm{h}}

\newcommand{\tb}{\bm{t}}

\newcommand{\yb}{\bm{y}}    %% bold y vector of observations

\newcommand{\zb}{\bm{z}}
\newcommand{\Omegab}{\bm{\Omega}} %%vector x domain

\newcommand{\zetab}{\bm{\zeta}}

\newcommand{\GP}{\mathcal{GP}}  %% GP distribution
\newcommand{\cN}{\mathcal{N}}   %% Gaussian 
\newcommand*\diff{\mathop{}\!\textrm{d}}

\newtheorem*{lemma*}{Lemma}

\newtheorem{theorem}{Theorem}

\newtheorem{lemma}{Lemma}
\newtheorem{corollary}{Corollary}
\newtheorem{remark}{Remark}

\providecommand{\keywords}[1]
{
  \small	
  \textbf{\textit{Keywords---}} #1
}
\title{A Fast and Calibrated Computer Model Emulator: An Empirical Bayes Approach
}
\author{Vojtech Kejzlar$^{1}$, Mookyong Son$^{2}$, Shrijita Bhattacharya$^{2}$, and Tapabrata Maiti$^{2}$\\
\small $^{1}$Department of Mathematics and Statistics, Skidmore College\\
\small $^{2}$Department of Statistics and Probability, Michigan State University}

\date{}
\begin{document}
\maketitle
\begin{abstract}
Mathematical models implemented on a computer have become the driving force behind the acceleration of the cycle of scientific processes. This is because computer models are typically much faster and economical to run than physical experiments. In this work, we develop an empirical Bayes approach to predictions of physical quantities using a computer model, where we assume that the computer model under consideration needs to be calibrated and is computationally expensive. We propose a Gaussian process emulator and a Gaussian process model for the systematic discrepancy between the computer model and the underlying physical process. This allows for closed-form and easy-to-compute predictions given by a conditional distribution induced by the Gaussian processes. We provide a rigorous theoretical justification of the proposed approach by establishing posterior consistency of the estimated physical process. The computational efficiency of the methods is demonstrated in an extensive simulation study and a real data example. The newly established approach makes enhanced use of computer models both from practical and theoretical standpoints.

\end{abstract}


%\tableofcontents

\section{Introduction}\label{sec:introduction}

With the advancements of computer architectures in the $21^{\text{th}}$ century, mathematical models implemented on a computer (\textit{computer models}) heavily contributed to the rapid speed-up of the cycle of  scientific processes. This is because computer models are generally much faster and economical to run than physical experiments. For instance, experiments conducted in high-energy particle colliders require budgets in billions of dollars and multinational collaborations. Additionally, many experiments related to natural events such as extreme weather phenomena, including tropical cyclones or tornadoes, are practically impossible to conduct.

Computer models, despite being an invaluable component of the process of scientific discovery, are imperfect representation of physical systems with each model evaluation often taking many hours. In this paper, we present an empirical Bayes approach for fast and statistically principled predictions of physical quantities using imperfect computer models that need to be calibrated with experimental observations. We particularly aim at those scenarios where computer models under consideration are complex and computationally too expensive to be used directly for predictions with quantified uncertainties. Our approach builds on the framework for computer model aided inference developed by \cite{KoH} that establishes the connection between experimental observations, computer model, and the systematic discrepancy (error) between the model and the physical process. The systematic discrepancy is modeled nonparametrically using a Gaussian process (GP) and the computer model is replaced by an emulator based also on a GP. This framework has reached high popularity over the past two decades with many applications in nuclear physics \citep{HigdonJPG15, King19}, climatology \citep{Sexton2012, Pollard2016}, and engineering \citep{williams2006, Plumlee16, Zhang2019}. There have been also various extensions of the original framework from both methodological and computational perspective. For example, \cite{Higdon08} consider computer models with high-dimensional output. \cite{Plumlee17} and \cite{MenGu18} study specific GP modeling choices to improve the predictive accuracy of the framework. \cite{kejzlar2020variational} develop variational inference based approach for approximation of posterior densities. \cite{Wu1}, \cite{Plumlee2019}, and lately \cite{ProjectCal2020} show theoretical properties of the framework under some modifications.

Despite these efforts, some of the practical challenges for computer enabled predictions with GPs remain. First, implementation of the framework \cite{KoH}  is never straightforward and typically requires considerable effort and experience, especially under some of the extensions listed in the previous paragraph. Second, a fully Bayesian approach becomes quickly computationally demanding with the increasing sample size, model complexity, and number of parameters. Third, in the absence of correct prior distributions, the full Bayesian models could be sensitive to the choice of hyperparameter values.   To avoid these complications, we consider an empirical Bayes approach, which can be viewed as an approximation to the fully Bayesian treatment. This approximation principle is well established for standard statistical models. We validate this in the context of calibrated computer models. Following are the specific contributions of this work:
\begin{itemize}
    \item[a)] Our methodology utilizes the statistical properties of GPs to establish easy-to-implement, closed-form, and fast-to-compute predictions of physical quantities using computationally expensive computer models that are calibrated with experimental observations. This includes a proposal of two estimators for plug-in model parameters with negligible loss of uncertainty on predictions that can be readily obtained using standard numerical solvers.
    \item[b)] We offer a fresh perspective on the framework of \cite{KoH} and provide its equivalent representation as a hierarchical model. As a consequence, we derive new theoretical properties of this framework and show that our proposed methodology estimates the values of underlying physical process consistently. Our theoretical analysis is based on an original extension of Schwartz's theorem for nonparametric regression problems with GP priors and an unknown but consistently estimated variance.
    \item[c)] We provide an extensive simulation study and demonstrate the computational efficiency of the proposed methodology compared with the Metropolis-Hasting algorithm (fully Bayesian implementation). We also conduct a sensitivity study of the fully Bayesian solution to prior selection and show that our methodology is preferred in the absence of proper and meaningful prior distributions. Additionally, we illustrate the opportunities provided by our method on an analysis of experimental nuclear binding energies. A fully documented Python code with our algorithm and examples is available at \url{https://github.com/kejzlarv/EB_Calibration}.
\end{itemize}

\subsection{Outline of this paper}\label{sec:introduction:outline}
In Section \ref{sec:framework}, we review the general framework for Bayesian inference with computer models. Section \ref{sec:Estimation} defines two plug-in estimators for GP model parameters and a consistent estimator of a noise variance component. Then, in Section \ref{sec:Consistency}, we discuss the theoretical properties of our approach and establish its statistical consistency. Section \ref{sec:applications} contains a simulation study that validates the methodology in this paper empirically. A real data application is also included in Section \ref{sec:applications}.

\section{Bayesian model for inference with computer models}\label{sec:framework}
Let us consider observations $\yb = (y_1, \dots, y_n)$ of a physical process $\zeta(\tb)$ depending on a known set of inputs $\tb_i,\; i=1,\cdots, n$ taking values in a compact and convex set $\Omegab \subset \mathbb{R}^p$, $p \ge 1$, following the relationship
\begin{equation}\label{eqn:physical_process}
    y_i = \zeta(\tb_i) + \sigma \epsilon_i, \quad i = 1, \dots, n,
\end{equation}
where $\sigma$ represents the scale of observational error, typically $\epsilon_i \stackrel{i.i.d.}{\sim} \cN(0,1)$. Our aim is to establish statistically principled predictions $\yb^* = (y^*_1, \dots, y^*_J)$ of the physical process $\zeta$ at new, yet to be observed, inputs $(\tb_1^*, \dots, \tb_J^*)$ using $\yb$ and a computer model $f_m$ defined as a mapping $(\tb, \thetab)\mapsto f_m(\tb, \thetab)$. As we can see, the computer model depends on an additional set of inputs $\thetab \in \Theta \subset \mathbb{R}^q$ that we call calibration parameters. These are considered fixed but unknown quantities common to all the observations $y_i$ and all the instances of the physical process that we intend to predict using calibrated computer model. The calibration parameters represent inherent properties of the physical process that cannot be directly measured or controlled in an experiment. In the most rudimentary form, one can think of the calibration parameters as parameters in standard regression problems. To this extent, we suppose the relationship between the observations $y$, physical process $\zeta$, and the computer model $f_m$ as proposed by \cite{KoH}:
\begin{equation} \label{eqn:complete_model}
	y_i = f_m(\tb_i,\thetab) + \delta(\tb_i) + \sigma \epsilon_i,
\end{equation}
where $\delta(\tb_i)$ represents an unknown systematic error between the computer model and the physical process. While $\delta(\tb_i)$ is intrinsically deterministic, a nonparametric approach using a GP prior model is typically imposed for Bayesian inference.

GPs are a convenient way of placing a distribution over a space of functions. By definition, we say that $\delta(\tb)$ has a GP distribution, if for every $i = 1, 2, 3 \dots$ the joint distribution of $\delta(\tb_1), \dots \delta(\tb_i)$ is multivariate normal. It is fully described by its mean and covariance functions that characterizes the relationship of the process at different inputs.

Typically, the mean function is chosen to be zero or some dense family of basis functions (wavelets, Fourier, polynomials) across the input domain:
\begin{equation}\label{eqn:mean_GP}
    m_{\delta}(\cdot) = \hb(\cdot)^T \betab,
\end{equation}
where $\hb(\cdot) = (h_1(\cdot), \dots h_r(\cdot))$ are the basis functions and $\betab$ is a \textit{hyperparameter}. A typical choice for  the covariance function is a stationary covariance function that depends on the inputs through $\tb - \tb'$. For example, a Gaussian kernel covariance function (also called squared exponential or radial basis function kernel) takes the form
\begin{equation}\label{eqn:rbf}
    k_{\delta}(\tb, \tb') = \eta \exp{\bigg(-\frac{1}{2}(\tb - \tb')^TM(\tb - \tb') \bigg)},
\end{equation}
where $M$ corresponds to a positive definite diagonal matrix of hyperparameters. We refer to the case of $M= \frac{1}{\ell^2}I$, for some $\ell > 0$, as an \textit{isotropic} version of the kernel, because it is invariant to the rotation. The case of $M$ with different diagonal terms is called an \textit{anisotropic} version of the kernel. Other popular choices for stationary covariance functions are Mat\'ern kernels, polynomial kernels, or exponential kernels \citep{RasmussenWilliams}.

It is important to note that one first needs to provide an estimate of the unknown parameter $\thetab$ according to the relationship \eqref{eqn:complete_model}, before making any predictions. The process of estimation of such parameter is called model calibration. In Bayesian sense, it corresponds to obtaining a full posterior distribution of $\thetab$ given data. Unfortunately, the calibration parameter $\thetab$ is non-identifiable in general.
Several authors have pointed this out 
and proposed various methods to mitigate the problem including \citep{Vlid,Ha,Plumlee17,Wu1,Wu2}. Our main goal here, nonetheless, is not the correct identification of $\thetab$, but a prediction. Thus the problem can be thought of as a ``black-box" based prediction such as the prediction based on neural networks or deep networks where parameters are part of the nonparametric models.

It is often the case that the evaluation of computer model $f_m$ is too expensive in terms of both time and space (memory). Common practice is to reduce the number of necessary computer model evaluations by considering a GP prior model. We use the following notation:
\begin{equation*}
			f_m(\tb, \thetab) \sim \mathcal{GP}(m_f(\tb, \thetab), k_f((\tb,\thetab),(\tb', \thetab'))).
\end{equation*}
In this setup, the data also include set of model evaluations $\zb = (z_1,\dots, z_{s})$ over a grid $\{(\widetilde{\tb}_1, \widetilde{\thetab}_1), \dots, (\widetilde{\tb}_{s}, \widetilde{\thetab}_{s})\}$. These are usually selected sequentially using some space-filling design such us uniform or Latin hypercube design \citep{MORRIS1995381}, which is a design that has a good coverage of the space with evenly distributed points in each one-dimensional projection. The complete dataset $\db$ in the case of computationally expensive models consists of $n$ observations $y_i$ from the physical process $\zeta$ and $s$ evaluations $z_j$ of the computer model $f_m$, i.e. $\bm{d} = (d_1, \dots, d_{n+s}) := (\yb,\zb) $. We shall denote the set of unknown parameters as $(\thetab, \phib, \sigma)$ with $\phib \in \Xi \subset \mathbb{R}^{q'}$ denoting the set of hyperparameters of GPs' mean and covariance functions. Consequently, the distribution of the complete dataset $\db$ conditioned on $(\thetab, \phib, \sigma)$ is

\begin{equation} \label{eqn:likelihod_d}
\bm{d}|\thetab, \phib, \sigma \sim N(M(\thetab,\phib), K(\thetab,\phib, \sigma)),
\end{equation}
where
\begin{equation} \label{eqn:likelihood_d_mean}
M(\thetab,\phib) = 
\begin{pmatrix}
M^{}_f(T_y(\thetab))  + M^{}_\delta(T_y)\\
M^{}_f(T_z(\widetilde{\thetab}))
\end{pmatrix},
\end{equation}
$M^{}_f(T_y(\thetab))$ is a column vector with $j^{\text{th}}$ element $m_f(\tb_j, \thetab)$, $M^{}_\delta(T_y)$ is a column vector with $j^{\text{th}}$ element $m_\delta(\tb_j)$, and $M^{}_f(T_z(\widetilde{\thetab}))$ is a column vector with $j^{\text{th}}$ element $m_f(\widetilde{\tb}_j, \widetilde{\thetab}_j)$. The covariance matrix of the multivariate normal distribution \eqref{eqn:likelihod_d} is
\small
\begin{align} \label{eqn:likelihood_d_covariance}
\begin{split}
&K(\thetab,\phib, \sigma) = \begin{pmatrix}
K^{}_f(T_y(\thetab), T_y(\thetab)) + K^{}_\delta(T_y, T_y) + \sigma^2I_n & K^{}_f( T_y(\thetab), T_z(\widetilde{\thetab}))\\
K^{}_f(T_z(\widetilde{\thetab}), T_y(\thetab)) & K^{}_f(T_z(\widetilde{\thetab}), T_z(\widetilde{\thetab})) 
\end{pmatrix}.
\end{split}
\end{align}
\normalsize
Here $K^{}_f(T_y(\thetab), T_y(\thetab))$ is the matrix with $(i,j)$ element $k_f((\tb_i,\thetab),(\tb_{j}, \thetab))$, $K^{}_\delta(T_y, T_y)$ is the matrix with $(i,j)$ element $k_\delta(\tb_i,\tb_{j})$, and  $K^{}_f(T_z(\widetilde{\thetab}), T_z(\widetilde{\thetab}))$ is the matrix with $(i,j)$ element $k_f((\widetilde{\tb}_i, \widetilde{\thetab}_i),(\widetilde{\tb}_j, \widetilde{\thetab}_j))$. We can define the matrix $K^{}_f( T_y(\thetab), T_z(\widetilde{\thetab}))$ similarly with the kernel $k_f$.

Under a fully Bayesian treatment, the predictions of $\yb^*$ are specified by the posterior predictive distribution $p(\yb^*|\db)$. It is obtained by integrating the conditional density  $p(\yb^*| \db, \thetab, \phib, \sigma)$, which is a multivariate normal density given by the statistical model \eqref{eqn:physical_process} and the specification of GPs, against the posterior density $p(\thetab, \phib, \sigma|\db)$. Analogical relationship holds for the predictions of new realizations of the physical process $\zetab^*$. The posterior density $p(\thetab, \phib, \sigma|\db)$, however, does not have a closed-form in general and one needs to resort to either Markov chain Monte Carlo (MCMC) methods for approximation or use variational techniques. This can be a non-trivial task to implement and requires some practical experience. Additionally, the nature of the marginal likelihood $p(\bm{d}|\thetab, \phib, \sigma)$ makes the problem harder to scale due to the complex structure of the covariance matrix $K(\thetab, \phib, \sigma)$, see \cite{KoH} and \cite{kejzlar2020variational} for further discussion.

To avoid these difficulties, we propose an empirical Bayes approach which instead of placing a (prior) distribution on $(\thetab, \phib, \sigma)$ estimates these parameters directly form the data. One can therefore utilize the convenience of GPs to obtain closed-form, simple, and fast predictions given by the conditional distribution $p(\yb^*| \db, \thetab, \phib, \sigma)$ (or $p(\zetab^*| \db, \thetab, \phib, \sigma)$). The proposed approach can be conceptualized as an approximation of the fully Bayesian treatment that neglects some of the uncertainty associated with the unknown parameters. 

\section{Prediction and parameter estimation}\label{sec:Estimation}
One of the main benefits of the empirical Bayes approach is that once we estimate the unknown parameters $(\thetab, \phib, \sigma)$, we can obtain a closed-form predictive distribution given these estimates. The framework additionally yields a principled approach for the inference of physical process $\zeta$ that is statistically consistent (shown below in Section \ref{sec:Consistency}). 

Here we formally derive the algorithm for prediction of physical quantities. Let us consider a set of new inputs $(\tb_1^*, \dots, \tb_J^*)$ at which we want to obtain prediction according to the model \eqref{eqn:complete_model}. The joint normality between $\db$ and $\yb^*$ implies that the conditional distribution $p(\yb^*|\db, \thetab, \phib, \sigma)$ is a multivariate normal distribution with the mean vector

\small
\begin{align}\label{eqn:VBI:predictiveY:mean}
\begin{split}
&M_{y^*}(\thetab, \phib, \sigma) = M^{}_f(T^*_y(\thetab))  + M^{}_\delta(T^*_y) + C_{*} K(\thetab, \phib, \sigma)^{-1}(\db - M(\thetab, \phib)),
\end{split}
\end{align}
\normalsize
and the covariance matrix

\small
\begin{align}\label{eqn:VBI:predictiveY:Cov}
\begin{split}
&K_{y^*}(\thetab, \phib, \sigma) = K^{}_f(T^*_y(\thetab), T^*_y(\thetab)) + K^{}_\delta(T^*_y, T^*_y) + \sigma^2I_J  - C_* K(\thetab, \phib, \sigma)^{-1}C^T_{*},
\end{split}
\end{align}
\normalsize
where

\small
\begin{align}\label{eqn:VBI:predictiveY:C}
\begin{split}
    &C_* = \begin{pmatrix}
K^{}_f(T^*_y(\thetab), T_y(\thetab)) + K^{}_\delta(T^*_y, T_y)\hspace{0.2cm} & K^{}_f(T^*_y(\thetab), T_z(\widetilde{\thetab}))
\end{pmatrix},
\end{split}
\end{align}
\normalsize
$M(\thetab, \phib)$ and $K(\thetab, \phib, \sigma)$ is the mean vector and the covariance matrix of the data likelihood $p(\db|\thetab, \phib, \sigma)$, $K^{}_f(T^*_y(\thetab), T^*_y(\thetab))$ is the matrix with $(i,j)$ element being $k_f((\tb^*_i,\thetab),(\tb^*_{j}, \thetab))$, $K^{}_f(T^*_y(\thetab), T_y(\thetab))$ is the matrix with $(i,j)$ element being $k_f((\tb^*_i,\thetab),(\tb_{j}, \thetab))$, $K^{}_\delta(T^*_y, T^*_y)$ is the matrix with $(i,j)$ element $k_\delta(\tb^*_i,\tb^*_{j})$, and $K^{}_\delta(T^*_y, T_y)$ is the matrix with $(i,j)$ element $k_\delta(\tb^*_i,\tb_{j})$. We can similarly define the matrix $K^{}_f( T^*_y(\thetab), T_z(\widetilde{\thetab}))$ with the kernel $k_f$ and the mean vectors $M^{}_f(T^*_y(\thetab))$ and $M^{}_\delta(T^*_y)$ as in the case of the likelihood \eqref{eqn:likelihod_d}. Analogical relationship holds for the conditional distribution of the new realizations from the physical process $p(\zetab^*|\db, \thetab, \phib, \sigma)$, where the mean vector $M_{\zeta^*}(\thetab, \phib, \sigma)$ is identical with \eqref{eqn:VBI:predictiveY:mean},
and the covariance matrix is

\small
\begin{align}\label{eqn:VBI:predictiveZ:Cov}
\begin{split}
&K_{\zeta^*}(\thetab, \phib, \sigma) = K^{}_f(T^*_y(\thetab), T^*_y(\thetab)) + K^{}_\delta(T^*_y, T^*_y) - C_{*} K(\thetab, \phib, \sigma)^{-1}C^T_{*},
\end{split}
\end{align}
\normalsize

The Algorithm \ref{alg:Pseudo} summarizes the procedure for predictions of physical quantities using imperfect and computationally expensive computer models.
\begin{algorithm}[h]
\DontPrintSemicolon
	\caption{Empirical Bayes algorithm for predictions of physical quantities using computer models\label{alg:Pseudo}}
		\KwIn{Data $\bm{d} = (\yb, \zb)$, mean and covariance functions for GPs, and new inputs $(\tb_1^*, \dots, \tb_J^*)$.}
		Use the experimental observations $\yb$ to compute the estimate of noise scale $\hat{\sigma}_n$\;
		Use $\db$ to obtain the estimates of GPs' hyperparameters $(\hat{\thetab}_{n+s}, \hat{\phib}_{n+s})$\;
		Compute $M_{y^*}(\hat{\thetab}_{n+s}, \hat{\phib}_{n+s}, \hat{\sigma}_n)$ and $K_{y^*}(\hat{\thetab}_{n+s}, \hat{\phib}_{n+s}, \hat{\sigma}_n)$ or $M_{\zeta^*}(\hat{\thetab}_{n+s}, \hat{\phib}_{n+s}, \hat{\sigma}_n)$ and $K_{\zeta^*}(\hat{\thetab}_{n+s}, \hat{\phib}_{n+s}, \hat{\sigma}_n)$ respectively to get the posterior predictive distribution\;
\end{algorithm}

\subsection{Parameter estimation}
As we have all closed-form expressions for the conditional distributions in Algorithm \ref{alg:Pseudo}, the computation avoids Monte Carlo sampling, hence negligible time is required compared to the sampling based approximations. This is assuming plugged-in parameter estimates.

To this extent, we propose the following estimator of the noise scale:
\begin{equation}\label{eqn:Sigma}
\hat{\sigma}_n =\sqrt{\frac{\sum_{i = 1}^{n-1}(y_{i + 1} - y_i)^2}{2(n - 1)}},
\end{equation}
where $y_i$ are the observations from the physical process under the model \eqref{eqn:physical_process}. The advantage of considering $\hat{\sigma}_n$ of this form is twofold. First, the estimator requires minimal computational effort. Second, $\hat{\sigma}_n$ is in fact a strongly consistent estimator (see Corollary \ref{col:Sigma} in Section \ref{sec:Consistency}) which turns out to be a crucial assumption for the theoretical validation of the empirical Bayes framework conducted in the following section.

\subsection{Estimation of hyperparameters}\label{subsec:framework_fitting}

\paragraph{Marginal data likelihood}
We first consider estimates of $(\thetab,\phib)$ as minimizers of a loss function that is reminiscent of the standard maximum likelihood approach, namely
\begin{equation}\label{eqn:loss_mle}
    L_{MLE}(\thetab, \phib) = - \log  p(\bm{d}|\thetab, \phib, \hat{\sigma}_n),
\end{equation}{}
with the negative log-likelihood being
\begin{align*}\label{loss_mle_likelihood}
&- \log  p(\bm{d}|\thetab, \phib, \hat{\sigma}_n) = \frac{1}{2} (\bm{d} - M(\thetab,\phib))^T K(\thetab,\phib, \hat{\sigma}_n) (\bm{d} - M(\thetab,\phib)) \\
&\qquad \qquad + \frac{1}{2} log | K(\thetab,\phib, \hat{\sigma}_n) | + \frac{n + s}{2} \log 2 \pi.
\end{align*}
We can interpret the minimizer of $L_{MLE}$ as a trade-off between the data-fit given by $\frac{1}{2} (\bm{d} - M(\thetab,\phib))^T K(\thetab,\phib, \hat{\sigma}_n) (\bm{d} - M(\thetab,\phib))$ and the model complexity penalty given by $\frac{1}{2} log | K(\thetab,\phib, \hat{\sigma}_n) |$  that depends only on model parameters and the variable inputs.

\paragraph{Predictive likelihood with K-fold cross-validation}
Another viable approach of estimating the parameters $(\thetab, \phib)$ is to base these on a model's predictive performance on unseen data. Cross-validation is a popular and robust approach to estimate this predictive performance that has been utilized across many statistical applications. See \cite{GPCVNIPS,RasmussenWilliams, CVIEEE} for applications with Gaussian processes. Here, we consider a K-fold cross-validation where the basic idea is to randomly partition the training dataset into $K$ subsets of roughly equal size. We then select $K-1$ subsets for training and consider the remaining set as a proxy for estimating the predictive performance. This is then repeated until we exhaust all the $K$ subsets for the purpose of validation with typical choices for $K$ being $3$, $5$, $10$, or $n$ (leave-one-out cross-validation).

Formally, let $\yb_i$ represent the $i^{th}$ subset of the observations $\yb$ and $\yb_{-i} = \yb \smallsetminus \yb_i$. The negative predictive log-likelihood under the K-fold cross-validation is
\begin{equation}\label{eqn:loss_cv}
    L_{CV(K)}(\thetab, \phib) = - \sum_{i}^{K} \log  p(\yb_i| \yb_{-i}, \zb, \thetab, \phib, \hat{\sigma}_n),
\end{equation}
The cross-validation should be more robust against  the model miss-specification and overfitting \citep{Wahba1990}.

\section{Theoretical analysis and posterior consistency}\label{sec:Consistency}
Below we represent the Bayesian model described in Section \ref{sec:framework} hierarchically using a set of prior distributions for a systematic exploitation of conjugacy. This representation of the model is crucial for the theoretical results obtained in Section \ref{sec:Consistency:Posterior}. It reframes the Bayesian model as a version of a nonparametric regression problem with a GP prior for $\zeta(\tb)$ and an additive noise. Namely, we define the model for data $\bm{d} = (d_1, \dots, d_{n+s}) = (\yb,\zb)$: 
\begin{align*}
    y_i &= \zeta(\tb_i) + \sigma \epsilon_i \quad \quad i=1,\dots,n,\\
    z_j &= f_m(\widetilde{\tb}_j, \widetilde{\thetab}_j), \quad \quad j=1,\dots,s,\\
    \epsilon_i &\stackrel{i.i.d.}{\sim} N(0, \sigma^2),
\end{align*}
where $z_j$'s are the realizations of computer model $f_m(\tb, \thetab)$ at pre-selected design points $(\widetilde{\tb}_j, \widetilde{\thetab}_j)$, and $y_i$'s are the experimental observations from the underlying physical process. Additionally, we consider the following GP priors:
\begin{align*}
\zeta(\tb)| f_m(\tb, \thetab), \delta(\tb) &\sim  f_m(\tb, \thetab) +  \delta(\tb), \\
\delta(\tb) &\sim \GP_\delta(m_{\delta}(\tb), k_{\delta}(\tb,\tb')), \\
    f_m(\tb, \thetab) &\sim \mathcal{GP}_f(m_f(\tb, \thetab), k_f((\tb,\thetab),(\tb', \thetab'))).
\end{align*}
Under this model, the conditional likelihoods for $y_i$ and $z_j$ are
\begin{align}
    p(y_i| \zeta(\tb_i), \sigma) &= \frac{1}{\sigma \sqrt{2 \pi}} \exp{\bigg( -\frac{(y_i - \zeta(\tb_i))^2}{2\sigma^2} \bigg)}, \\
    p(z_j| f_m(\widetilde{\tb}_j, \widetilde{\thetab}_j)) &= 1_{z_j = f_m(\widetilde{\tb}_j, \widetilde{\thetab}_j)} (z_j),
\end{align}
where $ p(z_j| f_m(\widetilde{\tb}_j, \widetilde{\thetab}_j))$ is a likelihood with the point mass at $z_j = f_m(\widetilde{\tb}_j, \widetilde{\thetab}_j)$. Consequently, the equivalence of the hierarchical formulation here and the model described in Section \ref{sec:framework} is given through the equality between the likelihood \eqref{eqn:likelihod_d} and the following integral, which shows that both model representations yield the same (marginal) data likelihood.
\begin{align*}\label{eqn:marginal}
    &\int_{\zetab} \int_{\tilde{f}_m} p(\zetab,\tilde{f}_m, \bm{d}|\thetab, \phib, \sigma) \diff \tilde{f}_m \diff \zetab = \\
    &\int_{\zetab} \int_{\tilde{f}_m} p(\bm{d}|\zetab , \tilde{f}_m, \thetab, \phib, \sigma) p(\zetab, \tilde{f}_m| \thetab, \phib) \diff \tilde{f}_m \diff \zetab  =\\
    & \int_{\zetab} \int_{\tilde{f}_m} \prod_i^{n} p(y_i| \zeta_i, \sigma) \prod_j^s  p(z_j| \tilde{f}_{m,j})p(\zetab, \tilde{f}_m| \thetab, \phib) \diff \tilde{f}_m \diff \zetab = \\
    & \int_{\zetab} \prod_i^{n} p(y_i| \zeta_i, \sigma) p(\zetab, \zb| \thetab, \phib) \diff{\zetab},
\end{align*}
where $\zetab = (\zeta(t_1), \dots,\zeta(t_n)) = (\zeta_1, \dots,\zeta_n)$
and $\tilde{f}_m = (f_m(\widetilde{\tb}_1, \widetilde{\thetab}_1), \dots, f_m(\widetilde{\tb}_s, \widetilde{\thetab}_s))$. The likelihood $p(\zetab, \zb| \thetab, \phib)$ is the multivariate normal distribution with the mean $M(\thetab,\phib)$ (see \eqref{eqn:likelihood_d_mean})
and the covariance
\begin{align*}
&K_p(\thetab,\phib) = \begin{pmatrix}
K^{}_f(T_y(\thetab), T_y(\thetab)) + K^{}_\delta(T_y, T_y) & K^{}_f( T_y(\thetab), T_z(\widetilde{\thetab}))\\
K^{}_f(T_z(\widetilde{\thetab}), T_y(\thetab)) & K^{}_f(T_z(\widetilde{\thetab}), T_z(\widetilde{\thetab}))
\end{pmatrix}.
\end{align*}
We leave the details of the integral computation for Appendix \ref{sec:Appendix:hierarchy}. Using this equivalent representation, we can gain a further insight into the role of the set of model runs $\zb$. Let us consider a function space $\mathcal{F}$ and a subset $\widetilde{\mathcal{F}} \subset \mathcal{F}$, then
\begin{equation}\label{eqn:Zrole}
p(\zeta \in \widetilde{\mathcal{F}}| \db, \thetab, \phib, \sigma) \propto \int_{\widetilde{\mathcal{F}}} \prod_i^{n} p(y_i| \zeta_i, \sigma) p(\zetab|\zb, \thetab, \phib) \diff{\zetab}.
\end{equation}
One can therefore interpret the model runs $\zb$ as an additional information provided by the computer model $f_m$ that enhances the GP prior distribution 
$p(\zetab|\zb, \thetab, \phib)$ over the physical process $\zeta$,  having the mean function
\begin{align}\label{eqn:zetaGP:mean}
\begin{split}
   &m_{\zeta}(\tb) = m_f(\tb, \thetab) + m_{\delta}(\tb) \\
   &+  \sum_{i,j=1}^s \kappa_{j,i} \Big[k_f((\tb,\thetab),(\widetilde{\tb}_j, \widetilde{\thetab}_j))\Big]\Big[z_i - m_f(\widetilde{\tb}_i, \widetilde{\thetab}_i)\Big],
   \end{split}
\end{align}
and the covariance function
\begin{align}\label{eqn:zetaGP:cov}
\begin{split}
   &k_{\zeta}(\tb,\tb') = k_f((\tb,\thetab),(\tb', \thetab)) + k_{\delta}(\tb,\tb') \\
   &- \sum_{i,j = 1}^s \kappa_{j,i} \Big[k_f((\tb,\thetab),(\widetilde{\tb}_j, \widetilde{\thetab}_j))\Big]\Big[k_f((\widetilde{\tb}_i, \widetilde{\thetab}_i),(\tb', \thetab))\Big],
\end{split}
\end{align}
where $\kappa_{j,i}$ is the $(j,i)$ element of the inverse matrix $K^{}_f(T_z(\widetilde{\thetab}), T_z(\widetilde{\thetab}))^{-1}$.

\subsection{Posterior consistency}\label{sec:Consistency:Posterior}
The revealing consequence of the previous discussion is that the \cite{KoH} framework is equivalent to the nonparametric regression model of an unknown function $\zeta(\tb)$ with the prior distribution $p(\zetab|\zb, \thetab, \phib)$. This is not only a new perspective on the popular framework, but also happens to be the key step that allows us to validate our empirical Bayes approach theoretically and establish the posterior consistency of the physical process when the prior $p(\zetab|\zb, \thetab, \phib)$ satisfies certain properties. To this end, rather than considering parametric forms of covariance kernels, the following results assume appropriate minimal smoothness of the GP prior over $\zeta$. This additionally means that any kernels with a smoothness parameter (e.g. Mat\'ern kernels) are considered to have the parameter fixed. Since the empirical Bayes estimator of smoothness parameter is not part of our procedure, the optimality of posterior concentration rate cannot be guaranteed. However, the focus of our asymptotic analysis is not on contraction rates but on consistency. We discuss the concrete examples of kernel functions that are sufficiently smooth at the end of this section.

In what follows, we suppose that the true underlying physical process $\zeta_0$ is a continuously differentiable function on the compact and convex set $\Omegab \subset \mathbb{R}^p$. Without loss of generality, we take $\Omegab = [0,1]^p$. Finally, we shall assume the plug-in estimates of the hyperparameters $(\hat{\thetab}_{n+s}, \hat{\phib}_{n+s})$ take values in some compact subset $\Upsilon \subset \Thetab \times \Xi$. This is a mild general condition that is satisfied by the hyperparameter estimators in Section \ref{sec:Estimation}, as long as the minimization of loss functions is constrained within some compact set. Analogous conditions have been considered recently by \cite{Teckentrup2020} in a GP regression setting similar to this paper. For any $\nu >0$, we aim to establish, under suitable conditions, the following:

\small
\begin{equation}\label{eqn:EB:statementFirstMention}
    p(\zeta \in W^C_{\nu,n}|y_1, \dots, y_n, \zb, \hat{\thetab}_{n+s}, \hat{\phib}_{n+s}, \hat{\sigma}_n) \xrightarrow[\text{n}]{\quad \quad} 0 \quad \text{a.s. } P_0,
\end{equation}
\normalsize
where $P_0$ denotes the joint conditional distribution of $\{y_i\}_{i = 1}^\infty$ given the true $\zeta_0$ and the true noise scale $\sigma_0$, $\hat{\sigma}_n$ is a strongly consistent estimator of $\sigma_0$, and 
\begin{equation}\label{eqn:EB:Wdefinition}
    W_{\nu, n}  = \bigg\{\zeta:\int|\zeta(\tb) - \zeta_0(\tb)| \diff Q_n(\tb) \leq \nu \bigg\},
\end{equation}
with $Q_n$ being the empirical measure on the design points given as $Q_n(\tb) = n^{-1} \sum_{i =1}^{n} \mathbbm{1}_{\tb_i}(\tb)$.

In Theorem \ref{thrm:consistency:1}, we first present a general result on the consistency of nonparametric regression problems and subsequently discuss the theorem's conditions in the context of the model described in Section
\ref{sec:framework}. This is based on the extensions of Schwartz's theorem for independent but non-identically distributed random variables given by \cite{CHOI2007b} and \cite{CHOI20071969}, where the authors assume $\sigma$ is included in $W_{\nu, n}$, and the posterior consistency is derived jointly for $\zeta$ and $\sigma$. On the other hand, the consistency of $\zeta$ conditioned on $\hat{\sigma}_n$, as stated in \eqref{eqn:EB:statementFirstMention}, requires a non-trivial modification of their original results. The proof of Theorem \ref{thrm:consistency:1} is provided in Appendix \ref{sec:Appendix:consistency1}.

\begin{theorem}\label{thrm:consistency:1}
Let $\{y_i\}_{i = 1}^\infty$ be independently and normally distributed with mean $\zeta(\tb_i)$ and standard deviation $\sigma$ with respect to a common $\sigma$-finite measure, where $\zeta$ belongs to a space of continuously differentiable functions on $[0,1]^p$ denoted as $\mathcal{F}$, and $\sigma > 0$. Let $\zeta_0 \in \mathcal{F}$ and let $P_0$ denote the joint conditional distribution of $\{y_i\}_{i = 1}^\infty$ given true $\zeta_0$ and $\sigma_0$. Let $\{U_n\}_{n=1}^\infty$ be a sequence of subsets of $\mathcal{F}$. Let $\zeta$ have a prior $\Pi(\cdot|\thetab, \phib)$ where $(\thetab, \phib)$ take values in a compact set $\Upsilon$. Then, under assumptions (A1)--(A3) (provided in Section \ref{sec:Consistency:Posterior:assumptions} below),

\begin{equation*}\label{eqn:statementTheorem}
    \sup_{(\thetab, \phib) \in \Upsilon} p(\zeta \in U_n^C|y_1, \dots, y_n, \thetab, \phib, \hat{\sigma}_n) \xrightarrow[\text{n}]{\quad \quad} 0 \quad \text{a.s. } P_0.
\end{equation*}
\end{theorem}

For the purpose of generality of Theorem \ref{thrm:consistency:1}, we do not explicitly condition on the set of model runs $\zb$. It is clear from our previous discussions (see \eqref{eqn:Zrole} in particular) that the model runs play the role of fixed constants in the prior distribution over $\zeta$. The dependence on $\zb$ in \eqref{eqn:EB:statementFirstMention} arises by setting $\Pi(\zetab|\thetab, \phib) := p(\zetab|\zb, \thetab, \phib)$, which is the GP prior distribution with the mean function \eqref{eqn:zetaGP:mean} and the covariance function \eqref{eqn:zetaGP:cov}.

\subsubsection{Assumptions for Theorem \ref{thrm:consistency:1}}\label{sec:Consistency:Posterior:assumptions}

As a matter of convenience, for any $0<\epsilon<1$ and $\zeta_0(\tb_i) = \zeta_{0,i}$ define
\begin{align*}
    \Lambda_i(\zeta_0,\zeta) &= \log \frac{p(y_{i}|\zeta_{0,i}, \sigma_0)}{p(y_{i}|\zeta_i, \sigma_0(1-\epsilon))},\\
    K_i(\zeta_0,\zeta) &= \mathbb{E}_{\zeta_0,\sigma_0}(\Lambda_i(\zeta_0,\zeta)), \\
    V_i(\zeta_0,\zeta) & = \mathbb{V}ar_{\zeta_0,\sigma_0}(\Lambda_i(\zeta_0,\zeta)).
\end{align*}

The following paragraph lists all the necessary conditions of Theorem \ref{thrm:consistency:1}:
\begin{itemize}
    \item[(A1)] Suppose there exists a set $B$ with $\Pi(B|\thetab, \phib) > 0$ for any $(\thetab, \phib) \in \Upsilon$, and for any $\Delta > 0$ a constant $0<\tilde{\epsilon}_1<1$, so that for any $\epsilon <\tilde{\epsilon}_1$:
\begin{itemize}
    \item[(i)] $\sum_{i = 1}^{\infty} \frac{V_i(\zeta_0,\zeta)}{i^2} < \infty$,  $\forall \zeta \in B$,
    \item[(ii)] $\Pi(B \cap \{\zeta: K_i(\zeta_0,\zeta) < \Delta\ \text{ for all  }i\}|\thetab, \phib) > 0$.
\end{itemize}

\item[(A2)] Suppose there exist tests $\{\Phi_{n}\}_{n=1}^{\infty}$, sets $\{\mathcal{F}_{n}\}_{n=1}^{\infty}$ and constants $C_{2},C_{1},c_{1} >0$ and $0<\tilde{\epsilon}_2 < 1$ so that:
\begin{itemize}
    \item[(i)] $\sum_{n=1}^{\infty}\mathbb{E}_{\zetab_{0},\sigma_0}\Phi_{n} < \infty$
    \item[(ii)] $\sup_{(\thetab, \phib) \in \Upsilon} \Pi(\mathcal{F}^C_n|\thetab, \phib) < C_1 e^{-c_1 n}$
    \item[(iii)]  There exists a constant $c_2 > 0$ such that for any $0<\epsilon < \tilde{\epsilon}_2$ the inequality $c_2 + \log (1-\epsilon) - \log(1+\epsilon) > 0$ holds and
    \begin{equation*}
        \sup_{\zeta \in U_{n}^C \cap \mathcal{F}_n} \mathbb{E}_{\zetab,\sigma_{0}(1+\epsilon)}( 1-\Phi_{n}) \leq C_2e^{-c_{2}n}.
    \end{equation*}
\end{itemize}

\item[(A3)] $\hat{\sigma}_n$ is strongly consistent, i.e $\hat{\sigma}_n \xrightarrow[\text{n}]{\quad \quad} \sigma_0 \quad \text{a.s. } P_0$.
\end{itemize}

We now discuss (A1)--(A3) in the context of the model described in Section \ref{sec:framework}. These fall into three general categories; the first one addresses prior positivity conditions ((A1) and (ii) of (A2)), second category is related to the existence of test functions $\Phi_n$ ((i) and (ii) of (A2)), and the last condition (A3) requires strong consistency of the noise scale estimator.

To verify conditions (A1) of Theorem \ref{thrm:consistency:1} for prior distributions, it is sufficient to show that the GP prior for $\zeta$ assigns positive probability to the following set for any $\omega >0$:
\begin{equation}\label{eqn:EB:consistency:delta}
    B_\omega = \left\{\zeta: \parallel \zeta - \zeta_0 \parallel_\infty < \omega \right\},
\end{equation}
where $\parallel \cdot \parallel _{\infty}$ denotes the supremum norm. For any $0<\epsilon<1$, a short calculation leads to

\begin{align*}\label{eqn:K:general}
    K_i(\zeta_0, \zeta)  =& \log(1-\epsilon) - \frac{1}{2}\left(1 - \frac{1}{(1-\epsilon)^2} \right) + \frac{[\zeta_0(\tb_i) - \zeta(\tb)]^2}{2 \sigma_0^2 (1-\epsilon)^2} \leq\\
    &\log(1-\epsilon) - \frac{1}{2}\left(1 - \frac{1}{(1-\epsilon)^2} \right) + \frac{\parallel\zeta_0 - \zeta\parallel_\infty^2}{2 \sigma_0^2 (1-\epsilon)^2}.
\end{align*}
Let $a(\epsilon) = \log(1-\epsilon) -1/2 + 1 / [2(1-\epsilon)^2]$, it is easy to see that $a(\epsilon)$ is positive and continuous at $\epsilon = 0$. Therefore, for every $\Delta > 0$, there exist $\omega > 0$ and $0<\tilde{\epsilon}_1<1$ so that $K_i(\zeta_0, \zeta) < \Delta$ for all $i$ and any $\epsilon < \tilde{\epsilon}_1$.

Additionally, for any $\epsilon < \tilde{\epsilon}_1$ and any $\omega > 0$
\begin{align*}
    V_i(\zeta_0, \zeta) &= \frac{1}{2}\left[\frac{1}{(1 - \epsilon)^2} - 1 \right]^2 + \left[\frac{[\zeta_0(\tb_i) - \zeta(\tb)]}{(1 - \epsilon)^2}\right]^2\\
    & < \infty \quad\text{uniformly in }i,
\end{align*}
and as a result, for all $\zeta \in B_\omega$, $\sum_{i = 1}^{\infty} \frac{V_i(\zeta_0, \zeta)}{i^2} < \infty$.  The prior condition (ii) of (A2) for the sieve $\mathcal{F}_n$ \eqref{eqn:EB:sieve} is addressed in Lemma \ref{lemma:consistency:prior}, see Appendix \ref{sec:Appendix:lemma} for proof.
\begin{lemma}\label{lemma:consistency:prior}
Let the mean function $m_{\zeta}(\cdot)$ of the GP prior for $\zeta$ defined on $[0,1]^p$ be continuously differentiable, and the covariance function $k_{\zeta}(\cdot,\cdot)$ has mixed partial derivatives up to order 4 that are continuous. Define,
\begin{align*}
    \rho^2_0(\thetab, \phib) &= \sup_{\tb \in [0,1]^p} \mathbb{V}ar \left(\zeta(\tb)|\zb, \thetab, \phib\right),\\
    \rho^2_i(\thetab, \phib) &= \sup_{\tb \in [0,1]^p} \mathbb{V}ar \left(\frac{\partial}{\partial t_i}\zeta(\tb)\bigg | \zb, \thetab, \phib\right), \quad  i = 1, \dots, p.
\end{align*}
Suppose that $\rho^2_i$ are continuous functions of $(\thetab, \phib)$ for all $(\thetab, \phib) \in \Upsilon$, $i = 0, \dots,p$, for any compact set $\Upsilon$. Then there exist constants $C, c > 0$ such that
\begin{equation*}
    \sup_{(\thetab, \phib) \in \Upsilon} p(\mathcal{F}^C_n|\zb, \thetab, \phib) < C e^{-c n},
\end{equation*}
where $\mathcal{F}_n$ are the sieves defined in \eqref{eqn:EB:sieve}.
\end{lemma}

Our approach to establish the existence of test functions $\{\Phi_n\}_{n=1}^{\infty}$ that satisfy the conditions (i) and (iii) in Theorem \ref{thrm:consistency:1} is similar to that of Theorem 2 in \cite{CHOI20071969}. We consider a sieve $\mathcal{F}_n$ which grows to the space of continuously differentiable functions on $[0,1]^p$. Namely, let
\begin{equation}\label{eqn:EB:sieve}
\begin{split}
    &\mathcal{F}_{n} =\left\{ \zeta:\; \parallel \zeta \parallel _{\infty}<M_{n}, \;\; \parallel \frac{\partial}{\partial t_{i}}\zeta \parallel _{\infty}<M_{n}, \;\;i=1,\cdots,p \right\},
    \end{split}
\end{equation}
where $M_n = \mathcal{O}(n^\alpha)$ for some $\alpha \in (\frac{1}{2},1)$. Each test is defined as a combination of tests over finitely many elements in the covering of $\mathcal{F}_n$. The existence of tests in the case of $W_{n, \nu}$ is given in Lemma \ref{thrm:tests} with  proof in Appendix \ref{sec:Appendix:tests}.

\begin{lemma}\label{thrm:tests}
Let $\mathcal{F}_n$ be the sieves defined in \eqref{eqn:EB:sieve}. For any $\nu > 0$ there exist tests $\{\Phi_n\}_{n=1}^{\infty}$ and constants $C$ and $0<\tilde{\epsilon}<1$ so that:
\begin{itemize}
    \item[(i)] $\sum_{n=1}^{\infty}\mathbb{E}_{\zetab_{0},\sigma_0}\Phi_{n} < \infty$
    \item[(ii)] There exists a constant $c> 0$ such that for any $0<\epsilon < \tilde{\epsilon}$ the inequality $c + \log (1-\epsilon) - \log(1+\epsilon) > 0$ holds and
    \begin{equation*}
        \sup_{\zeta \in W_{n,\nu}^C \cap \mathcal{F}_n} \mathbb{E}_{\zetab,\sigma_{0}(1+\epsilon)}( 1-\Phi_{n}) \leq Ce^{-cn}.
    \end{equation*}
\end{itemize}
\end{lemma}

As we have suggested in Section \ref{sec:Estimation}, the estimator $\hat{\sigma}_n$ defined in \eqref{eqn:Sigma} is in fact strongly consistent estimator of the true scale parameter $\sigma_0$.

\begin{theorem}\label{thrm:Sigma}
Suppose $\zeta_0(\tb)$ represents the true physical process and $\sigma_0^2$ be the true value of the experimental error variance, where $\tb \in \Omegab$ is a compact and convex subset of $\mathbb{R}^p$ and $\zeta_0$ is continuously differentiable on $\Omegab$. Let $P_0$ denote the joint conditional distribution of $\{y_i\}_{i = 1}^\infty$ given true $\zeta_0$ and $\sigma_0^2$. Also assume that the following holds about the design points $\tb_i$:
\begin{equation}\label{eqn:DesignReq}
\tag{AD}
    \sup_{i \in \{1, \dots, n\}, j \in \{1, \dots, p\}} |t_{i+1, j} - t_{i,j}| \xrightarrow[\text{n}]{\quad \quad} 0,
\end{equation}
then
\begin{equation}\label{eqn:ConsistenSigma}
\hat{\sigma}_n^2 \xrightarrow[\text{n}]{\quad \quad} \sigma_0^2 \quad \text{a.s. } P_0.
\end{equation}
\end{theorem}
The proof of Theorem \ref{thrm:Sigma} is given in Appendix \ref{sec:Appendix:Sigma}. The continuous mapping theorem directly implies the following.
\begin{corollary}\label{col:Sigma}
Under the assumptions of Theorem \ref{thrm:Sigma},
\begin{equation}\label{eqn:ConsistencySigmaCol}
   \hat{\sigma}_n = \sqrt{\hat{\sigma}_n^2} \xrightarrow[\text{n}]{\quad \quad} \sigma_0 \quad \text{a.s. } P_0.
\end{equation}
\end{corollary}
\begin{remark}\label{remark:Sigma}
The assumption \eqref{eqn:DesignReq} is satisfied by a design that contains at least one point in each hypercube $H$ in $\Omegab$ with its Lebesgue measure $\lambda(H) \ge \frac{1}{K n}$, for some constant $0 < K \le 1$. This is, for example, the case of equally spaced design.
\end{remark}

Below we present Theorem \ref{col:consistency} whose corollary is, under the additional assumption of  $(\hat{\thetab}_{n+s}, \hat{\phib}_{n+s})$ taking values in some compact set $\Upsilon$, the almost sure consistency result \eqref{eqn:EB:statementFirstMention}.

\begin{theorem}\label{col:consistency} 
Let $P_0$ denote the joint conditional distribution of $\{y_i\}_{i = 1}^\infty$ given true $\zeta_0$ and $\sigma_0$. Let $m_\zeta(\cdot)$ and $k_\zeta(\cdot, \cdot)$ be the mean and covariance functions of the GP prior for $\zeta$ satisfying the conditions of Lemma \ref{lemma:consistency:prior}. Assume that for any compact set $\Upsilon$ and any $\omega > 0$, $p(B_\omega|\zb, \thetab, \phib) > 0$, where $(\thetab, \phib) \in \Upsilon$. If $\hat{\sigma}_n$ is a strongly consistent estimator of $\sigma_0$, then for any $\nu > 0$
\small
\begin{equation}\label{eqn:EB:statementCol}
    \sup_{(\thetab, \phib) \in \Upsilon} p(\zeta \in W^C_{\nu,n}|y_1, \dots, y_n, \zb, \thetab, \phib, \hat{\sigma}_n) \xrightarrow[\text{n}]{\quad \quad} 0 \quad \text{a.s. } P_0.
\end{equation}
\normalsize
\end{theorem}

Theorem \ref{col:consistency} is a direct consequence of Lemmas \ref{lemma:consistency:prior} and \ref{thrm:tests}, and Theorem \ref{thrm:consistency:1}.

\paragraph{Prior conditions: concrete examples} The key sufficient condition for the convergence of empirical Bayes posterior \eqref{eqn:EB:statementFirstMention} is the prior positivity condition requiring $p(B_\omega|\zb, \thetab, \phib) > 0$ for any $\omega$ and $(\thetab, \phib) \in \Upsilon$ which was extensively studied by \cite{ghosal2006} and \cite{TOKDAR200734}. Specifically, Theorem 4 of \cite{ghosal2006} states that this condition is satisfied for a GP with continuous sample paths and continuous mean and covariance functions, as long as $\zeta_0$ and $m_\zeta$ belong to  reproducing kernel Hilbert space (RKHS) of the covariance function $k_\zeta$. First, the continuity of GP's sample paths is given by the application of Theorem 5 in \cite{ghosal2006} which requires the same smoothness conditions as Lemma \ref{lemma:consistency:prior} in this section. It should be clear from \eqref{eqn:zetaGP:mean} and \eqref{eqn:zetaGP:cov} that $m_\zeta$ is continuous on $[0,1]^p$, and $k_\zeta$ has continuous mixed partial derivatives up to $4^{th}$ order on $[0,1]^p$, as long as the same holds about $m_f$ and $m_\delta$ (commonly used mean functions including polynomials are analytic functions) and respectively $k_f$ and $k_\delta$. For example, the product of one-dimensional Mat\'ern kernels with fixed smoothness parameter $\lambda > 2$ (tensor-product Mat\'ern kernel) and the squared exponential kernel are sufficiently smooth \citep{williams2006}.  Second, \cite{TOKDAR200734} show that the RKHS of $k_\zeta$ spans the space of continuously differentiable functions on $[0,1]^p$, if $k_\zeta$ is a product of $p$ isotropic and integrable univariate covariance functions with continuous mixed partial derivatives up to order 4. The squared exponential kernel and the tensor-product Mat\'ern kernel with smoothness $\lambda > 2$ satisfy these requirements, including the continuity of $\rho^2_i$ for $i = 0, \dots, p$.

This, of course, does not directly imply that such choices for $k_f$ and $k_\delta$ result in the conditional covariance $k_\zeta$ whose RKHS spans the space of continuously differentiable functions. However, our numerical study show that with increasing number of computer model evaluations $s$ \--- obtained using a space filling design \--- the covariance function $k_{\delta}$ quickly dominates (see Appendix \ref{sec:Appendix:kernel} for details). Such behavior is not unexpected since the increasing number of runs $s$ effectively reduces the uncertainty about emulated computer model. This indicates that $k_{\zeta}$ and $k_{\delta}$ behave asymptotically same, with respect to the $s$. Additionally, the simulation study conducted in Section \ref{sec:applications} strongly suggests that choosing the squared exponential kernel leads to consistent predictions.

\section{Numerical analysis and applications}\label{sec:applications}
The main objective of this section is to establish the efficiency of the empirical Bayes method in Algorithm \ref{alg:Pseudo} and to support the consistency result presented in section \ref{sec:Consistency}. All this while sacrificing minimally in terms of the fidelity of UQ as compared to the fully Bayesian treatment. To this extent, we consider a simulation study where we compare our method (under both $L_{MLE}$ and $L_{CV(K)}$) to a fully Bayesian treatment with the posterior samples obtained using the standard Metropolis-Hastings algorithm \citep{BDA}. We also conduct a prior sensitivity analysis of the fully Bayesian treatment to further the practical advantages of the empirical Bayes. Finally, we demonstrate the opportunities provided by our method for science practitioners through predictions of nuclear binding energies using the Liquid Drop Model.

\subsection{Simulation study: Transverse harmonic wave}
Let us consider a simple computer model representing a periodic wave disturbance that moves through a medium and causes displacement of individual atoms or molecules in the medium. This is called a transverse harmonic wave, where the displacement $f_m((t, x), \thetab)$ of a particle at location $x$ over time $t$ is given by
\begin{equation}\label{eqn:wave}
    f_m((t, x), \thetab) = \theta_1 \sin \big(kx - \theta_2 t + \psi\big),
\end{equation}
where $\theta_1$ represents the amplitude of the wave, and $\theta_2$ is the frequency of the wave. The model also depends on the wave number $k$, which is reciprocal to the wave length, and the phase constant $\psi$. For the purpose of this example, we shall consider these to be known values with $k = 5$ and $\psi = 1$, and define the model inputs $(t,x)$ over the space $[0,1]^2$ (we assume that the length and time units are all equal to one). The true physical process is modeled according to
\begin{equation}\label{eqn:wave:process}
\zeta_0(t, x) = f_m((t,x), \thetab) + \delta(t, x) = \theta_1 \sin \big(5x - \theta_2 t + 1\big) + \beta,
\end{equation}
where $\beta = 1$ is a constant systematic error of the model and $\thetab = (\theta_1, \theta_2)$ are arbitrarily set to be (1.2, 1.8). 

\subsubsection{Data generation and design}\label{sec:sensiticity:design}
We generate the experimental observation according to the model \eqref{eqn:complete_model} with the true value of the observation error scale $\sigma_0 = 0.2$, where the model inputs $(t,x)$ are chosen using the Latin hypercube design over the full space $[0,1]^2$. The space filling properties of the design guarantee decreasing bias of the estimator $\hat{\sigma}_n$ with an increasing sample size. Additionally, we assume that the computer model for the periodic wave disturbance is computationally expensive and generate the set of model runs $\zb$ using, again, the Latin hypercube design, now over $[0,1]^2\times[0,2]^2$. In each of the subsequent scenarios, the amount of experimental observations is equal to the number of computer model runs, i.e. $n =s$. We define the GP priors for $f_m$ and $\delta$ to have zero means and the covariance functions

\small
\begin{align*}
    &k_f(\{t, x,\thetab\},\{t',x', \thetab'\}) = \eta_f \cdot \exp{(-\frac{\|t - t' \|^2}{2\ell^2_t} -\frac{\|x - x' \|^2}{2\ell^2_x} - \frac{\|\theta_1 - \theta_1' \|^2}{2\ell^2_{\theta_1}} - \frac{\|\theta_2 - \theta_2' \|^2}{2\ell^2_{\theta_2}})}, \\
   &k_\delta(\{t, x\}, \{t', x'\}) = \eta_\delta \cdot \exp{(-\frac{\|t - t' \|^2}{2\nu^2_t} -\frac{\|x - x' \|^2}{2\nu^2_x})}.
\end{align*}
\normalsize
The hyperparameters in this scenario are therefore $\phib =(\eta_f, \ell_t,\ell_x, \ell_{\theta_1}, \ell_{\theta_2}, \eta_\delta, \nu_t, \nu_x)$. For the fully Bayesian treatment, we choose inverse gamma priors with shape and scale parametrization for $(\sigma, \eta_f, \eta_\delta)$, gamma priors with shape and rate parametrization for the length scales, and independent normal distributions for the calibration parameters $(\theta_1, \theta_2)$. As we demonstrate below, the performance of the MCMC-based fit can vary greatly with different prior selections. To asses this effect, we consider the following prior variations: inverse gamma distributions with the shape fixed at 3 and the scale taking values in $\{0.5, 1, 2, 4, 8\}$, gamma distribution with the rate equal to 3 and the shape taking values in $\{1, 5\}$, and the normal distribution with the mean $\mu_{\theta} \in \{0,1, 1.5\}$ and the standard deviation $\sigma_\theta \in  \{0.25,0.5, 1, 2\}$. These choices reflect both fairly informative priors (e.g. $\mu_\theta = 1.5$ and $\sigma_\theta = 0.25$) and non-informative priors, given the spans of both the input space $[0,1]^2$ and the parameter space $[0,2]^2$.

\subsubsection{Results}

\begin{figure*}[!h]
	\begin{centering}
		\includegraphics[width=0.9\textwidth]{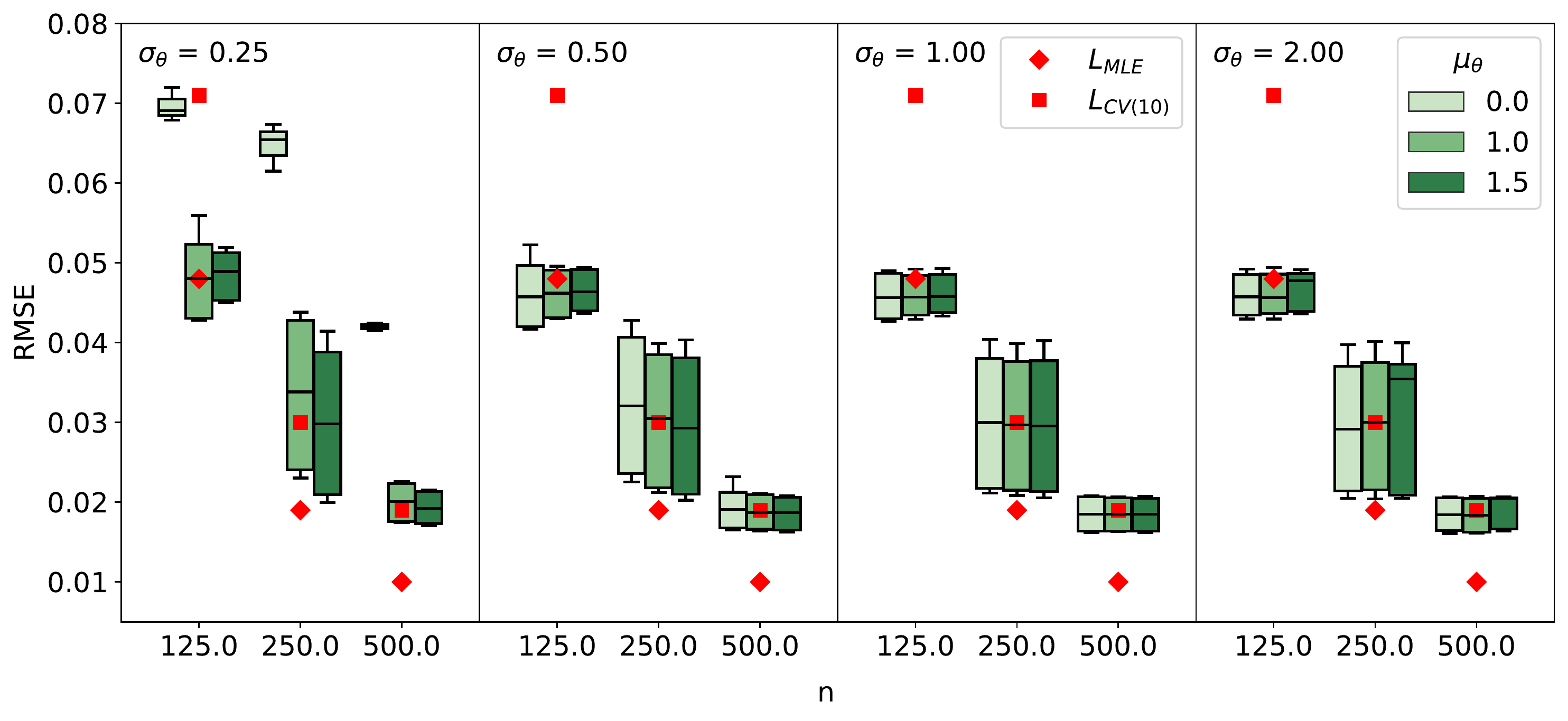}
		\caption{The RMSE of the empirical Bayes approach and the fully Bayesian treatment. The results are grouped according to the values of prior means $\mu_\theta$ and standard deviations $\sigma_\theta$ used in the Metropolis-Hastings algorithm. The box-plots represent the distribution of RMSE values obtained with the MCMC-based fits across the prior combinations described in Section \ref{sec:sensiticity:design}. The GP hyperparameters for the empirical Bayes approach were estimated using Algorithm \ref{alg:Pseudo}.}
		\label{fig:Sim1:1:RMSE}
	\end{centering}
\end{figure*}

Figure \ref{fig:Sim1:1:RMSE} shows the root mean squared errors (RMSEs) of predictions of new realizations from the true physical process \eqref{eqn:wave:process} evaluated on a testing datasets of 225 realizations over a uniform grid on $[0,1]^2$. The predictions are taken to be the posterior predicative means under each method. Each box-plot in Figure \ref{fig:Sim1:1:RMSE} represents the distribution of RMSEs obtained through the MCMC-based fits for given values of $\mu_\theta$ and $\sigma_\theta$. We consider the estimates of hyperparameters using the $L_{MLE}$ loss and the predictive likelihood loss function with 10-fold cross-validation under the empirical Bayes approach. The noise scale parameter was estimated using the consistent estimator $\hat{\sigma}_n$ defined in Section \ref{sec:Estimation}. 

In general, the proposed empirical Bayes approach performs comparably with the fully Bayesian treatment and monotonously decreases with the increasing size of the dataset. In particular, the RMSE under the $L_{CV(10)}$ loss is larger than the other methods for the smallest size of training dataset considered, however, the RMSE under the $L_{MLE}$ loss is the smallest for the larger training sets. The likely reason for the slightly better performance of the empirical Bayes is that the parameter estimates given by the minimization of $L_{MLE}$ and $L_{CV(10)}$ are purely data driven, whereas the fully Bayesian approach needs to account for prior uncertainties. This observation is consistent with the sensitivity of the predictions to the prior selection clearly visible in Figure \ref{fig:Sim1:1:RMSE}. A choice of strongly informative prior that is far from the underlying truth, such as $\mu_\theta = 0$ and $\sigma_\theta = 0.25$, can yield especially poor fit even for large training sets. Thus, in the absence of proper and meaningful prior distributions, an empirical Bayes approach may be preferable besides its other advantages as discussed in this article. Overall, the empirical Bayes fit can be readily obtained in several minutes using standard numerical solvers while sampling from posterior distributions can take hours.

\begin{figure*}[!h]
	\begin{centering}
		\includegraphics[width=0.9\textwidth]{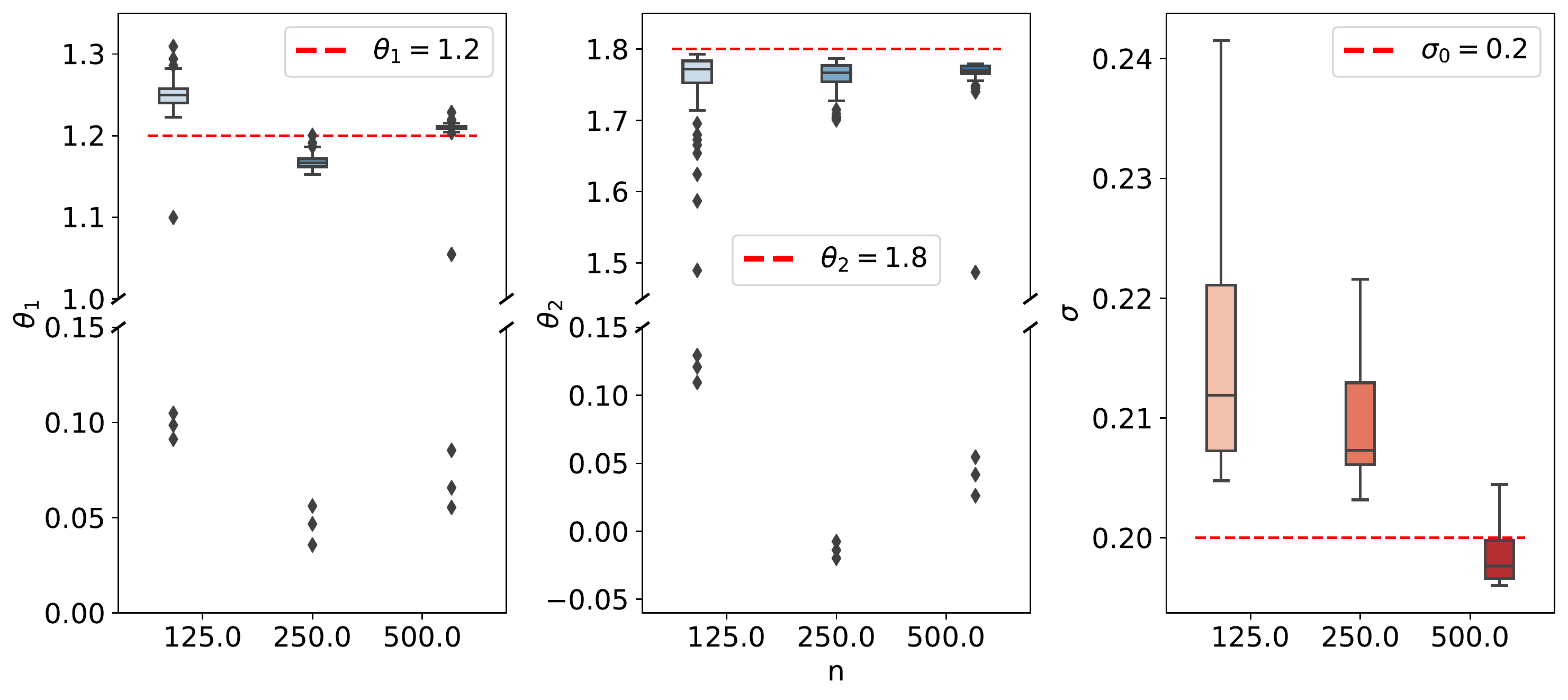}
		\caption{The distribution of posterior means of the calibration parameters and the noise scale obtained with the Metropolis-Hastings algorithm. Unlike in Figure \ref{fig:Sim1:1:RMSE}, the box-plots were aggregated over all the prior choices. The values used to generate the simulation data were $(\theta_1, \theta_2) = (1.2, 1.8)$ and $\sigma_0 = 0.2$.}
		\label{fig:Sim1:1:Params}
	\end{centering}
\end{figure*}

It took approximately 2 hours to obtain $10^4$ samples in the scenario with the largest sample size on a standard PC with 4 cores. For completeness, we also show the estimates of calibration parameters and the noise scale under each method in Figure \ref{fig:Sim1:1:Params} and Table \ref{table:Sim1:params}. Posterior means were taken as the estimates under the fully Bayesian solution. We can see a reasonable match between the approximate empirical Bayes method and the Metropolis-Hastings algorithm for many of the prior choices. The first notable difference is a series of outlying estimates of the calibration parameters under the MCMC-based fit. These are the consequence of the aforementioned strongly informative priors. The second difference is in terms of the noise scale estimate $\hat{\sigma}_n$. This is expected since the estimate is unbiased asymptotically.

\setlength{\tabcolsep}{3pt}
\renewcommand{\arraystretch}{0.7}
\begin{table}[H]
\centering
\begin{tabular}{{c|cc|cc|cc}}
		\hline 
		\hline \noalign{\smallskip}
		   & \multicolumn{2}{c|}{$n = 125$, $s=125$} & \multicolumn{2}{c|}{$n=250$, $s=250$} & \multicolumn{2}{c}{$n = 500$, $s=500$} \\
		  \noalign{\smallskip} \cline{2-7}\noalign{\smallskip}
 & $L_{MLE}$ & $L_{CV(10)}$ & $L_{MLE}$ & $L_{CV(10)}$ & $L_{MLE}$ & $L_{CV(10)}$  \\
\noalign{\smallskip} \cline{1-7} \noalign{\smallskip}
$\theta_1$ & 1.197 & 1.217 & 1.160 & 1.251  & 1.207 & 1.206 \\
$\theta_2$  & 1.781 & 1.787 & 1.805 & 1.799 & 1.792 & 1.818 \\
$\sigma$ & \multicolumn{2}{c|}{0.328} & \multicolumn{2}{c|}{0.259} & \multicolumn{2}{c}{0.228} \\\hline\hline \noalign{\smallskip}
	\end{tabular}
\caption{The estimates of calibration parameters and the noise scale under the empirical Bayes approach. The values used to generate the simulation data were $(\theta_1, \theta_2) = (1.2, 1.8)$ and $\sigma_0 = 0.2$. \label{table:Sim1:params}\label{table:Sim1:RMSE}}
\end{table}

\begin{figure*}
	\begin{centering}
		\includegraphics[width=0.95\textwidth]{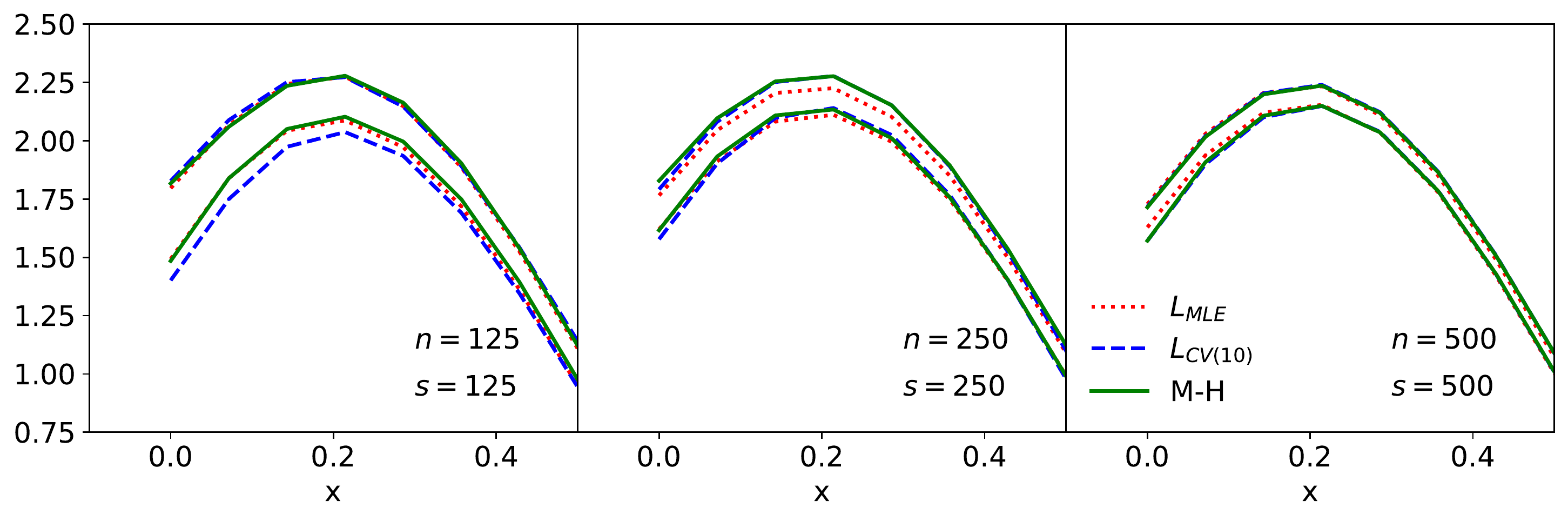}
		\caption{Details of $95 \%$ credible bands of posterior predictive distributions under the empirical Bayes approach and the fully Bayesian approach of Metropolis-Hastings algorithm. These were plotted at $t=0.21$.}
		\label{fig:Sim1:1:CB}
	\end{centering}
\end{figure*}

\begin{figure*}
	\begin{centering}
		\includegraphics[width=0.95\textwidth]{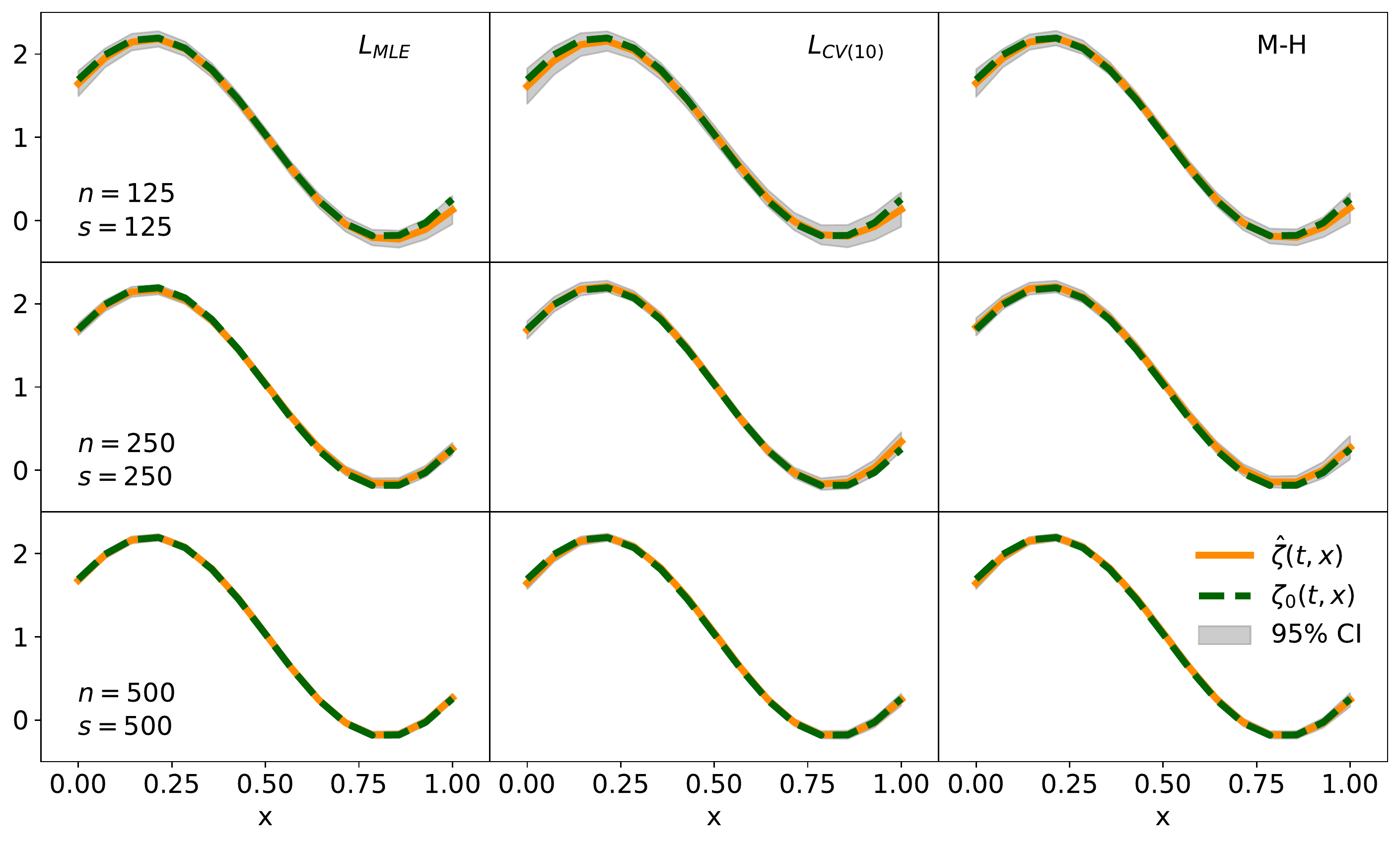}
		\caption{Comparison of the convergence to the true physical process $\zeta_0(t,x)$ under the empirical Bayes approach and the fully Bayesian implementation given by the Metropolis-Hastings algorithm. The dashed line represents the true process $\zeta_0$, and the solid line corresponds to the mean of posterior predictive distributions under respective method. The curves with $95 \%$ credible intervals (shaded area) are plotted at $t=0.21$.}
		\label{fig:Sim1:1}
	\end{centering}
\end{figure*}

Figure \ref{fig:Sim1:1:CB} and  Figure \ref{fig:Sim1:1} show the loss in terms of UQ is negligible under the empirical Bayes approach as compared to the fully Bayesian treatment for all practical purposes. For clarity, we display only the results of inverse gamma priors with shape $3$ and scale $1$, gamma priors with shape 1 and rate 3, and normal priors with mean 0 and standard deviation 2. These are fairly non-informative priors. We can see that the empirical Bayes approach slightly overestimates the uncertainty for smaller sample size, but this quickly diminishes as the sample size increases. This is likely the consequence of the inflation of the noise scale given by the bias of $\hat{\sigma}_n$ which diminishes with the increasing sample size as expected. See Appendix \ref{sec:Appendix:Sim1} for additional figures of the empirical Bayes fit at the time locations $t = 0$, $t = 0.43$, $t = 0.71$, and $t = 1$.

\subsection{Liquid Drop Model for nuclear binding energies}
Nuclear physics is one of many fields that has recently experienced a surge in the applications of Bayesian statistics due to its intuitive way to describe uncertainties probabilistically. GP modeling and its variants have been prominently used in the context of computationally expensive theoretical mass models for either emulation or modeling of systematic discrepancies to produce precise and quantified predictions of nuclear observables \citep{HigdonJPG15,Neufcourt18-1, Neufcourt18-2, Schnuck2020}.

To illustrate our framework for computer enabled predictions on a real data example, we shall consider the 4-parameter Liquid Drop Model (LDM) \citep{Myers1966,Kirson2008,Benzaid2020} of nuclear binding energy, which is the minimum energy needed to break the nucleus of an atom into free protons and neutrons. It is equivalent (energy-mass equivalence explained by $E=mc^2$) to the mass defect that corresponds to the difference between the mass number of a nucleus and its actual measured mass. This difference is caused by the energy released in the event of atom's creation. The LDM is a simple yet reasonably accurate description of the atomic nucleus given by the semi-empirical mass formula:
\begin{equation}\label{eqn:LDM}
\begin{split}
&E_{\rm B}(N,Z) =\theta_{\rm vol}A - \theta_{\rm surf}A^{2/3} - \theta_{\rm sym} \frac{(N-Z)^2}{A} - \theta_{\rm C} \frac{Z(Z-1)}{A^{1/3}}.
\end{split}
\end{equation}
The LDM is a function of the proton number $Z$ and the neutron number $N$ ($A = Z + N$ is the mass number) that depends on a set of calibration parameters $\thetab = (\theta_{\rm vol}, \theta_{\rm surf}, \theta_{\rm sym}, \theta_{\rm C})$. These have physical meaning that represent the volume, surface, symmetry and Coulomb energy (see \cite{krane1987introductory} for details). The semi-empirical mass formula is particularly suitable example, because it provides a good fit for heavy nuclei and somewhat poor fit for light nuclei. This clearly points to the existence of a systematic model discrepancy that is also supported in the literature \citep{Reinhard2006, Yuan2016, Kejzlar2020}.

\begin{figure}[H]
	\begin{centering}		
	\includegraphics[width=1\textwidth]{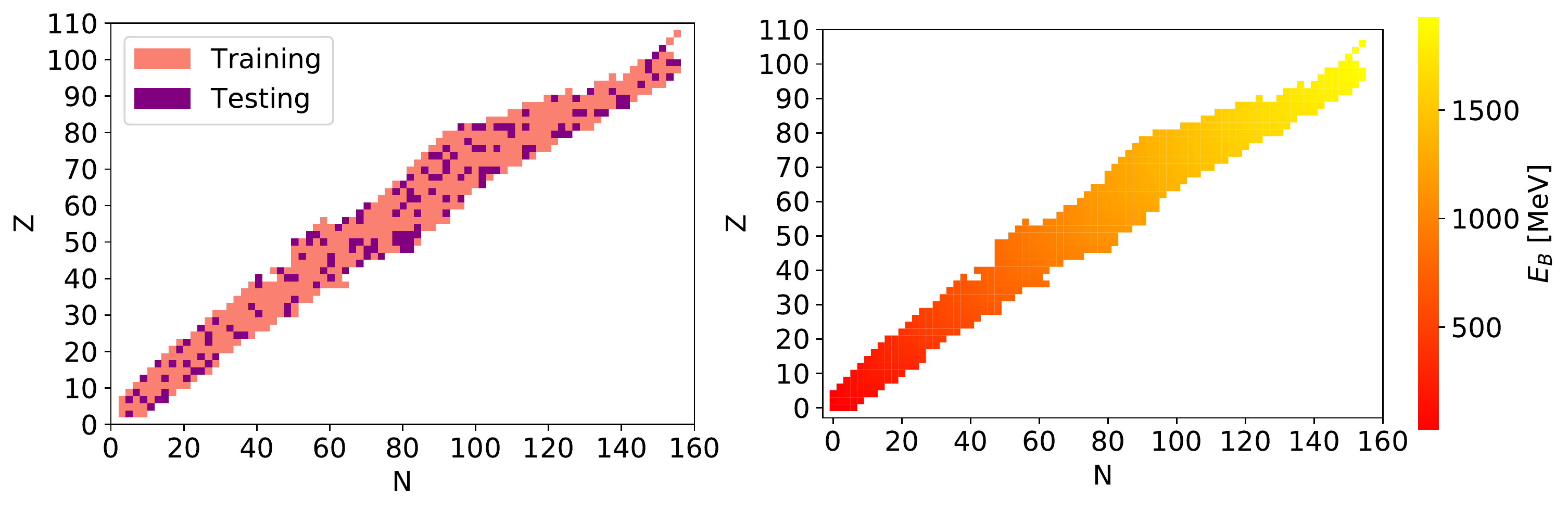}
		\caption{Binding energies of even-even nuclei in AME2003 dataset divided into the testing and  training datasets.}
		\label{fig:AME2003}
	\end{centering}
\end{figure}

We now present an analysis of 595 experimental binding energies of even-even nuclei from the AME2003 dataset \citep{AME2003} (publicly available at \url{http://amdc.impcas.ac.cn/web/masstab.html}) randomly divided into a training set of 450 nuclei and a testing set of 145 nuclei, see Figure \ref{fig:AME2003}.
We consider the statistical model \eqref{eqn:complete_model} and model the systematic discrepancy $\delta$ with zero mean GP and the isotropic squared exponential covariance function. For the purpose of this example, we also assume that the LDM is computationally expensive (or not directly accessible) and regard it is an unknown function of $(Z,N)$ and $\thetab$. Similarly to the discrepancy $\delta$, we assign a GP prior to $E_{\rm B}(N,Z)$ with zero mean and the isotropic squared exponential covariance function. To this extent, we additionally generated a set of 900 model evaluations using the Latin hypercube design over the space spanning all reasonable values of the parameters $\thetab$ as given by the nuclear physics literature \citep{Weizsacker1935,Bethe36,Myers1966,Kirson2008,Benzaid2020}. Corresponding nuclear configurations, the inputs $(Z, N)$, were randomly assigned to the generated values of $\thetab$ from a set of two times duplicated training nuclei. We also want to point out that this is not the first application of GP modeling in the context of the LDM. See \cite{Bertsch17} for instance. We conducted a similar study previously using a fully Bayesian approach with posterior distributions approximated through variational inference \cite{kejzlar2020variational}.

\subsubsection{Results}\label{sec:LDM:results}
The predictions of nuclear binding energies were computed as the means of the posterior predictive distribution \eqref{eqn:VBI:predictiveY:mean} conditioned on the estimates of the calibration parameters $\thetab$, GP's hyperparameters $\phib$, and the noise scale $\hat{\sigma}_n$. The estimates for $(\thetab, \phib)$ were obtained numerically as the minimizers of $L_{MLE}$ and $L_{CV(10)}$. The priors for the GP hyperparameters in the case of the fully Bayesian treatment are discussed in Appendix \ref{sec:app:LDM}.
\setlength{\tabcolsep}{4pt}
\renewcommand{\arraystretch}{0.9}
\begin{table}[H]
	\begin{tabular}{l|cccc|c}
		\hline 
		\hline \noalign{\smallskip} 
		 & \multicolumn{4}{c|}{Parameter estimates} & Testing error \\ \noalign{\smallskip} \cline{2-6}\noalign{\smallskip} 
		& $\theta_{\textrm{vol}}$ & $\theta_{\textrm{surf}}$ & $\theta_{\textrm{sym}}$ & $\theta_{\textrm{C}}$ & RMSE (MeV)\\\noalign{\smallskip}  \hline\noalign{\smallskip} 
		$L_{MLE}$ & 15.07 & 15.58 & 22.00 &  0.68 & 1.16 \\
		$L_{CV(10)}$ & 15.08  & 16.08 & 21.19 & 0.67 & 1.26 \\
		M-H & 15.32  & 16.09 & 22.09 & 0.70 & 1.16 \\
		\noalign{\smallskip} \hline\hline 
		%LS & 15.14  & 15.93 & 22.00 & 0.68 & 4.10 \\ \noalign{\smallskip} \hline\hline 
	\end{tabular}
	\caption{The RMSEs of the predictions evaluated on 145 even-even nuclei from the AME2003 dataset. The parameter estimates are also listed. The posterior means are shown in the case of the Metropolis-Hastings algorithm. \label{table:LDM_results}}
\end{table}

Table \ref{table:LDM_results} gives the RMSE values calculated on the testing set of 145 even-even nuclei for the empirical Bayes approach and also the Metropolis-Hastings algorithm. The calibration parameter estimates are also provided with values that do not significantly differ between the methods considered. The resulting RMSEs are $1.1-1.3$ MeV which is a consistent result with our previous study in \cite{kejzlar2020variational} that was conducted on a larger dataset, however, under a fully Bayesian stetting. Overall, this is quite a remarkable result given the considerable effort that needs to be put forth to implement the fully Bayesian solution. We were able to obtain the empirical Bayes predictions under 10 minutes using the standard optimization modules in Python, while the Metropolis-Hastings algorithm needed close to 8 hours to generate $1.5\times 10^4$ samples.

\section{Conclusion}\label{sec:conclusion}
We presented and studied an empirical Bayes approach to prediction of physical quantities using computer model, where we assumed that the computer model under consideration needs to be calibrated and is computationally too expensive to be used directly for inference. To this extent, we proposed a GP emulator and utilized the structural convenience of GPs to formulate closed-form and easy-to-compute predictions of new observations from a physical process. These predictions are obtained through conditional predictive distributions with plugged-in estimates of calibration parameters, GP hyperparameters, and experimental noise scale. A strongly consistent estimator for the noise scale and two sensible estimators for the remaining parameters (defined as minimizers of two alternative loss functions) were provided.

Theoretical study and justification of the proposed methodology were also given: we revisited hierarchical models and established an equivalent representation of the framework of \cite{KoH} as a nonparametric regression model with GP prior for an unknown function corresponding to the underlying physical process. Consequently, we derived a non-trivial extension of Schwartz's theorem for nonparametric regression problems. The application of this results shows that our method consistently estimates the underlying true physical process, assuming smoothness of the mean and covariance functions of GP priors and the existence of a strongly consistent estimator of the noise scale. To the best of our knowledge, this is the first such posterior consistency result under the original model of \cite{KoH}. Nonetheless, our theoretical study is by no means exhaustive. For example, the asymptotic analysis in this work does not focus on posterior concentration rates. The derivation of optimal minimax rates on the contraction of the posterior requires further extensively study on the RKHS properties of Gaussian process priors as in \cite{vandervaart2008}  and we thereby leave it as future work.  We also assumed that any covariance kernels with smoothness parameter have the parameter fixed. We refer the reader to \cite{Belitser2008, Flores2012, szabo2013, sniekers2015, Knapik2016, rousseau2017, serra2017}, for discussion about posterior consistency in non-parametric regression related problems with smoothness estimator.

A simulation study that empirically supports the consistency result was given in Section \ref{sec:applications}. The speed and efficiency of the empirical Bayes approach was demonstrated in comparison to the fully Bayesian approach of Metropolis-Hastings algorithm. Both methods yield comparable results in terms of UQ and quality of the predictions, however, the Metropolis-Hastings algorithm is significantly slower and its implementation requires considerable effort. Additionally, our sensitivity study strongly suggests that the empirical Bayes approach may be preferable in the absence of proper and meaningful prior distributions. Finally, to show the opportunities given by our method for practitioners, we analyzed a dataset of experimental binding energies using the Liquid Drop Model.

The general framework presented in this paper can be wived as a fast and computationally efficient approximation to the sampling based fully Bayesian approach for calibration of computer models that neglects some uncertainty of unknown parameters. Our empirical studies show that this loss becomes quickly negligible with the increasing size of datasets.

\section*{Acknowledgments}
	The authors thank the reviewers and the Editor for their helpful comments and ideas. The research is partially supported by National Science Foundation funding DMS1952856.

%\bibliographystyle{Chicago}    % basic style, author-year citations
%\bibliographystyle{spbasic} 
\bibliography{biblio}   % name your BibTeX data base
\newpage
\begin{appendices}

\section{Equivalency of hierarchical model} \label{sec:Appendix:hierarchy}
To establish the equivalency between the Bayesian model given by the data likelihood $p(\bm{d}|\thetab, \phib, \sigma)$ and the hierarchical model (see Section \ref{sec:Consistency}), we need to show that the following equality holds
\begin{equation}\label{eqn:app:equivalency}
   p(\bm{d}|\thetab, \phib, \sigma) =\int_{\zetab} \prod_i^{n} p(y_i| \zeta_i, \sigma) p(\zetab, \zb| \thetab, \phib) \diff {\zetab},
\end{equation}
where $\zetab = (\zeta(t_1), \dots,\zeta(t_n)) = (\zeta_1, \dots,\zeta_n)$ and the density $p(\zetab, \zb| \thetab, \phib)$ is the multivariate normal distribution with mean the mean $M(\thetab,\phib)$ (see \eqref{eqn:likelihood_d_mean})
and the covariance

\small
\begin{align*}
&K_p(\thetab,\phib) =\begin{pmatrix}
K^{}_f(T_y(\thetab), T_y(\thetab)) + K^{}_\delta(T_y, T_y) & K^{}_f( T_y(\thetab), T_z(\widetilde{\thetab}))\\
K^{}_f(T_z(\widetilde{\thetab}), T_y(\thetab)) & K^{}_f(T_z(\widetilde{\thetab}), T_z(\widetilde{\thetab}))
\end{pmatrix}
= \begin{pmatrix}
C_{11} & C_{12}\\
C_{21} & C_{22}
\end{pmatrix}.
\end{align*}
\normalsize

For the ease of notation, let us now assume $M(\thetab,\phib) = (M_{y}^T, M_{z}^T)^T$. Then

\small
\begin{align*}
&\int_{\zetab} \prod_i^{n} p(y_i| \zeta_i, \sigma) p(\zetab, \zb| \thetab, \phib) \diff{\zetab} \\
&= \int_{\zetab} \frac{1}{(2\pi)^{n/2} |\sigma^2 I_n |^{1/2}} \text{exp}\bigg( -\frac{1}{2} (\yb - \zetab)^T(\sigma^2I_n)^{-1}(\yb - \zetab)\bigg)\times \frac{1}{(2\pi)^{(n+s)/2} |K_p |^{1/2}} \\
&\quad \times \text{exp}\bigg(-\frac{1}{2} \binom{\zetab - M_{y}}{\zb - M_{z}}^T K_p^{-1}\binom{\zetab - M_{y}}{\zb - M_{z}}\bigg) \diff \zetab \\
&= \frac{1}{(2\pi)^{(n+s)/2} |K |^{1/2}} \text{exp}\bigg(-\frac{1}{2} \binom{\yb - M_{y}}{\zb - M_{z}}^T K^{-1}\binom{\yb - M_{y}}{\zb - M_{z}}\bigg) \times \int_{\zetab} \frac{|K |^{1/2}}{(2\pi)^{n/2} |\sigma^2 I_n|^{1/2} |K_p |^{1/2}} \\
&\quad \times \text{exp}\bigg(-\frac{1}{2}(\yb - \zetab)^T(\sigma^2I_n)^{-1}(\yb - \zetab)\bigg) \times \text{exp}\bigg(-\frac{1}{2} \binom{\zetab - M_{y}}{\zb - M_{z}}^T K_p^{-1}\binom{\zetab - M_{y}}{\zb - M_{z}}\bigg) \\
&\quad \times \text{exp}\bigg( \frac{1}{2} \binom{\yb - M_{y}}{\zb - M_{z}}^T K^{-1}\binom{\yb - M_{y}}{\zb - M_{z}}\bigg) \diff \zetab \\
&= \frac{1}{(2\pi)^{(n+s)/2} |K |^{1/2}} \text{exp}\bigg(-\frac{1}{2} \binom{\yb - M_{y}}{\zb - M_{z}}^T K^{-1}\binom{\yb - M_{y}}{\zb - M_{z}}\bigg).
\end{align*}
\normalsize
The integral is equal to 1 since it is an integration of multivariate normal probability density function over $\zeta$ with covariance $( (\sigma^2 I_n)^{-1} + (C_{11} - C_{12}C^{-1}_{22} C_{21})^{-1})^{-1}$. Namely

\begin{align*}
 \frac{|K |^{1/2}}{|\sigma^2 I_n|^{1/2} |K_p |^{1/2}}  &= \frac{|C_{22}|^{1/2} |C_{11} + \sigma^2 I_n  - C_{12}C^{-1}_{22} C_{21} |^{1/2}}{|\sigma^2 I_n|^{1/2} |C_{22}|^{1/2} |C_{11} - C_{12}C^{-1}_{22} C_{21} |^{1/2}} \\
 &= \frac{|C_{11} + \sigma^2 I_n  - C_{12}C^{-1}_{22} C_{21} |^{1/2}}{|\sigma^2 I_n|^{1/2} |C_{11} - C_{12}C^{-1}_{22} C_{21} |^{1/2}} \\
 &= \frac{|A + B|^{1/2}}{|A|^{1/2}|B|^{1/2}} = \frac{1}{|A|^{1/2}|B|^{1/2} |A + B|^{ -1/2}} \\
 &= \frac{1}{(|A^{-1}||B^{-1}| |A + B|)^{ -1/2}} = \frac{1}{|A^{-1}B^{-1}A + A^{-1}B^{-1}B|^{ -1/2}} \\
 &= \frac{1}{|A^{-1}B^{-1}A + A^{-1}|^{ -1/2}} = \frac{1}{|A^{-1}(B^{-1} + A^{-1})A|^{ -1/2}} \\
 &= \frac{1}{(|A^{-1}| |(B^{-1} + A^{-1})| |A|)^{ -1/2}}  = \frac{1}{|(B^{-1} + A^{-1})^{-1}|^{1/2}}
\end{align*}
where we used the Schur complement identity for determinants in the first equality and
\begin{align*}
A &=  C_{11} - C_{12}C^{-1}_{22} C_{21}, \\
B &= \sigma^2 I_n.
\end{align*}
Lastly, considering the notation
\begin{align*}
K_p^{-1}= 
\begin{pmatrix}
C^-_{11} & C^-_{12}\\
C^-_{21} & C^-_{22}
\end{pmatrix}
\end{align*}
we have
\small
\begin{align*}
&\exp{\bigg(-\frac{1}{2}(\yb - \zetab)^T(\sigma^2I_n)^{-1}(\yb - \zetab)\bigg)} \exp{ \bigg(-\frac{1}{2} \binom{\zetab- M_{y}}{\zb - M_{z}}^T K_p^{-1}\binom{\zetab - M_{y}}{\zb - M_{z}}\bigg)}
\\&\times \exp{\bigg(\frac{1}{2} \binom{\yb - M_{y}}{\zb - M_{z}}^T K^{-1}\binom{\yb - M_{y}}{\zb - M_{z}}\bigg)}  \propto \exp{\bigg(-\frac{1}{2}\zetab^T ((\sigma^2 I_n)^{-1}+  C^-_{11})\zetab +\zetab^T \bb \bigg)},
\end{align*}
\normalsize
where $C^-_{11} = C_{11} - C_{12}C^{-1}_{22} C_{21}$ and $\bm{b}$ is a constant column vector. This shows that integral is indeed equal to 1 as stated, and the equality \eqref{eqn:app:equivalency} holds.

\section{Proof of Theorem \ref{thrm:consistency:1}}\label{sec:Appendix:consistency1}
Note that for any $\epsilon > 0$, the posterior probability of interest $p(\zeta \in U_n^C|y_1, \dots, y_n, \thetab, \phib, \hat{\sigma}_n)$ can be bound from the above as
\begin{align*}
    &p(\zeta \in U_n^C|y_1, \dots, y_n, \thetab, \phib, \hat{\sigma}_n)\leq p(\zeta \in U_n^C|y_1, \dots, y_n, \thetab, \phib, \hat{\sigma}_n)1_{ \{\left|\frac{\hat{\sigma}_{n}}{\sigma_0}-1\right| \leq \epsilon\}} + 1_{ \{\left|\frac{\hat{\sigma}_{n}}{\sigma_0}-1\right| > \epsilon\}},
\end{align*}
\normalsize
where
\small
\begin{align*}
&p(\zeta \in U_n^C|y_1, \dots, y_n, \thetab, \phib, \hat{\sigma}_n)1_{ \{\left|\frac{\hat{\sigma}_{n}}{\sigma_0}-1\right| \leq \epsilon\}} \\ 
& \leq \Phi_{n} + \frac{(1 -\Phi_{n})\int_{U_n^c \cap \mathcal{F}_n} \prod_{i = 1}^n\frac{p(y_i|\zeta_i,\hat{\sigma}_{n})}{p(y_i|\zeta_{0,i},\sigma_0)} 1_{ \{\left|\frac{\hat{\sigma}_{n}}{\sigma_0}-1\right| \leq \epsilon\}} \diff \Pi(\zetab|\thetab, \phib)}{\int_{\mathcal{F}} \prod_{i = 1}^n\frac{p(y_i|\zeta_i,\hat{\sigma}_{n})}{p(y_i|\zeta_{0,i},\sigma_0)} \diff \Pi(\zetab|\thetab, \phib)}\\
&\quad + \frac{\int_{U_n^c \cap \mathcal{F}^C_n} \prod_{i = 1}^n\frac{p(y_i|\zeta_i,\hat{\sigma}_{n})}{p(y_i|\zeta_{0,i},\sigma_0)} 1_{ \{\left|\frac{\hat{\sigma}_{n}}{\sigma_0}-1\right| \leq \epsilon\}} \diff \Pi(\zetab|\thetab, \phib)}{\int_{\mathcal{F}} \prod_{i = 1}^n\frac{p(y_i|\zeta_i,\hat{\sigma}_{n})}{p(y_i|\zeta_{0,i},\sigma_0)} \diff \Pi(\zetab|\thetab, \phib)} \\
&= \Phi_{n} + \frac{\mathbf{I}_{1n}(y_1, \dots, y_n, \thetab, \phib, \hat{\sigma}_n, \epsilon)}{\mathbf{I}_{3n}(y_1, \dots, y_n, \thetab, \phib, \hat{\sigma}_n)} + \frac{\mathbf{I}_{2n}(y_1, \dots, y_n, \thetab, \phib, \hat{\sigma}_n, \epsilon)}{\mathbf{I}_{3n}(y_1, \dots, y_n, \thetab, \phib, \hat{\sigma}_n)}.
\end{align*}
\normalsize
Since the assumption (A3) implies that $1_{ \{\left|\frac{\hat{\sigma}_{n}}{\sigma_0}-1\right| > \epsilon\}}\xrightarrow[\text{n}]{\quad} 0 \text{ a.s. } P_0$, it is enough to show that there exists $\epsilon > 0$ so that
\begin{align}
&\sup_{(\thetab, \phib) \in \Upsilon} \Phi_{n} \xrightarrow[\text{n}]{\quad} 0 \text{ a.s. } P_0, \label{eq:1}\\
&\sup_{(\thetab, \phib) \in \Upsilon} e^{\beta_{1}n}\mathbf{I}_{1n}(y_1, \dots, y_n, \thetab, \phib, \hat{\sigma}_n, \epsilon) \xrightarrow[\text{n}]{\quad} 0 \text{ a.s. } P_0, \label{eq:2}\\
&\sup_{(\thetab, \phib) \in \Upsilon} e^{\beta_{2}n}\mathbf{I}_{2n}(y_1, \dots, y_n, \thetab, \phib, \hat{\sigma}_n, \epsilon) \xrightarrow[\text{n}]{\quad} 0 \text{ a.s. } P_0,\label{eq:3}\\
&\inf_{(\thetab, \phib) \in \Upsilon} e^{\beta_{3}n}\mathbf{I}_{3n}(y_1, \dots, y_n, \thetab, \phib, \hat{\sigma}_n) \xrightarrow[\text{n}]{\quad} \infty \text{ a.s. } P_0, \label{eq:4}
\end{align}
for some $\beta_1, \beta_2, \beta_3 > 0$ where $\beta_3 \leq \min\{\beta_1, \beta_2\}$.

The rest of the proof follows the general steps of the proof of Theorem 1 in \cite{CHOI20071969} and Theorem 9 in \cite{CHOI2007b} with some non-trivial treatment of the constant $\epsilon$. Similarly to \cite{CHOI2007b}, we assume that the covariance function is smooth enough so that the supremum of the variance of the Gaussian process is continuous with respect to ($\thetab, \phib$) on the compact set $\Upsilon$. We shall provide step by step details below.
\paragraph{Step 1)} By Markov inequality, for any $\rho > 0$
\begin{align*}
   \sum_{n=1}^{\infty}P_0(\Phi_{n} > \rho) \leq \frac{1}{\rho} \sum_{n=1}^{\infty}\mathbb{E}_{\zeta_{0},\sigma_0}\Phi_{n},
\end{align*}
which due to the condition (i) of (A2) and the first Borel-Cantelli Lemma yields
\begin{equation*}
    \Phi_{n} \xrightarrow[\text{n}]{\quad} 0 \text{ a.s. } P_0.
\end{equation*}
Since this does not depend on $(\thetab, \phib)$, it implies \eqref{eq:1}.
\paragraph{Step 2)} By Fubini's theorem and for any $0<\epsilon < \tilde{\epsilon}_2$
\small
\begin{align*}
    &\mathbb{E}_{\zeta_{0},\sigma_0}(\mathbf{I}_{1n}(y_1, \dots, y_n, \thetab, \phib, \hat{\sigma}_n, \epsilon))\\
    &= \mathbb{E}_{\zeta_{0},\sigma_0} \bigg[(1 -\Phi_{n})\int_{U_n^c \cap \mathcal{F}_n} \prod_{i = 1}^n\frac{p(y_i|\zeta_i,\hat{\sigma}_{n})}{p(y_i|\zeta_{0,i},\sigma_0)} 1_{ \{\left|\frac{\hat{\sigma}_{n}}{\sigma_0}-1\right| \leq \epsilon\}} \diff \Pi(\zetab|\thetab, \phib)\bigg] \\
    &=\int_{U_n^c \cap \mathcal{F}_n} \int (1 -\Phi_{n}) \prod_{i = 1}^n\frac{p(y_i|\zeta_i,\hat{\sigma}_{n})}{p(y_i|\zeta_{0,i},\sigma_0)} 1_{ \{\left|\frac{\hat{\sigma}_{n}}{\sigma_0}-1\right| \leq \epsilon\}}\diff P_0 \diff \Pi(\zetab|\thetab, \phib) \\
    &\leq {\left(\frac{\sigma_{0}(1-\epsilon)}{\sigma_{0}(1+\epsilon)}\right)}^{-n} \int_{{U_{n}}^{C}\cap \mathcal{F}_{n}} \mathbb{E}_{\zetab,\sigma_{0}(1+\epsilon)}  [(1-\Phi_{n})] \diff\Pi(\zetab|\thetab, \phib)  \\
     &\leq   {\left(\frac{1-\epsilon}{1+\epsilon}\right)}^{-n} \sup_{\zeta \in U_n^C \cap \mathcal{F}_n} \mathbb{E}_{\zeta,\sigma_{0}(1+\epsilon)}[( 1-\Phi_{n})] \\
     &\leq {\left(\frac{1-\epsilon}{1+\epsilon}\right)}^{-n} C_2 e^{-c_2 n} = C_2 e^{-\tilde{c}_\epsilon n},
\end{align*}
\normalsize
where $\tilde{c}_\epsilon = c_2 + \log(1-\epsilon) - \log(1+\epsilon)$ together with condition (iii) of (A2) implies $\tilde{c}_\epsilon > 0$. Thus
\begin{align*}
    P_0 \bigg\{\mathbf{I}_{1n}(y_1, \dots, y_n, \thetab, \phib, \hat{\sigma}_n, \epsilon) \ge e^{-\tilde{c}_\epsilon \frac{n}{2}} \bigg\} &\leq C_2 e^{\tilde{c}_\epsilon \frac{n}{2}}e^{-\tilde{c}_\epsilon n} \\
    &= C_2e^{-\tilde{c}_\epsilon \frac{n}{2}}.
\end{align*}
Therefore, for any $\epsilon > 0$ so that $\epsilon < \tilde{\epsilon}_2$ there exists a constant $\tilde{c}_\epsilon$ for which the first Borel-Cantelli Lemma implies
\begin{equation*}
    e^{\tilde{c}_\epsilon \frac{n}{4}}\mathbf{I}_{1n}(y_1, \dots, y_n, \thetab, \phib, \hat{\sigma}_n, \epsilon) \xrightarrow[\text{n}]{\quad} 0 \text{ a.s. } P_0.
\end{equation*}
Since this does not depend on $(\thetab, \phib)$, it implies \eqref{eq:2}.
\paragraph{Step 3)} If we proceed as in the step 2), the Fubini's theorem implies
\begin{align*}
&\mathbb{E}_{\zeta_{0},\sigma_0}(\mathbf{I}_{2n}(y_1, \dots, y_n, \thetab, \phib, \hat{\sigma}_n, \epsilon))\\
&= \mathbb{E}_{\zeta_{0},\sigma_0} \bigg[\int_{U_n^c \cap \mathcal{F}_n} \prod_{i = 1}^n\frac{p(y_i|\zeta_i,\hat{\sigma}_{n})}{p(y_i|\zeta_{0,i},\sigma_0)} 1_{ \{\left|\frac{\hat{\sigma}_{n}}{\sigma_0}-1\right| \leq \epsilon\}} \diff \Pi(\zetab|\thetab, \phib)\bigg] \\
&\leq {\left(\frac{\sigma_{0}(1-\epsilon)}{\sigma_{0}(1+\epsilon)}\right)}^{-n} \int_{{U_{n}}^{C}\cap \mathcal{F}_{n}^C} \mathbb{E}_{\zeta,\sigma_{0}(1+\epsilon)}  [1] \diff\Pi(\zetab|\thetab, \phib)\\
&\leq {\left(\frac{1-\epsilon}{1+\epsilon}\right)}^{-n} \Pi(\mathcal{F}^C_n|\thetab, \phib).
\end{align*}
The condition (ii) of (A2) and the first Borel-Cantelli Lemma implies that for any $\epsilon < \frac{1 - e^{-c_1}}{1 + e^{-c_1}}$:
\begin{equation*}
    \sup_{(\thetab, \phib) \in \Upsilon} e^{\tilde{k}_\epsilon \frac{n}{4}}\mathbf{I}_{2n}(y_1, \dots, y_n, \thetab, \phib, \hat{\sigma}_n, \epsilon) \xrightarrow[\text{n}]{\quad} 0 \text{ a.s. } P_0,
\end{equation*}
where $\tilde{k}_\epsilon = c_1 + \log(1 -\epsilon) - \log(1+\epsilon)$.
\paragraph{Step 4)} To prove \eqref{eq:4}, given any $0<\rho < 1$, we first observe the following:
\begin{align*}
&\mathbf{I}_{3n}(y_1, \dots, y_n, \thetab, \phib, \hat{\sigma}_n)\\
&\ge \mathbf{I}_{3n}(y_1, \dots, y_n, \thetab, \phib, \hat{\sigma}_n)1_{ \{\left|\frac{\hat{\sigma}_{n}}{\sigma_0}-1\right| \leq \rho\}} \\
& \ge {\left(\frac{1-\rho}{1+\rho}\right)}^{n} \int_{\mathcal{F}} \prod_{i = 1}^n\frac{p(y_i|\zeta_i,\sigma_0(1-\rho))}{p(y_i|\zeta_{0,i},\sigma_0)} \diff \Pi(\zetab|\thetab, \phib).
\end{align*}
Let us now define $\log_+(x)= \max\{0, \log (x)\}$ and $\log_-(x)= -\min\{0, \log (x)\}$ as well as
\begin{align*}
    W_{i}&=\log_{+} \frac{p(y_{i}|{\zeta_{0,i}},\sigma_{0})}{p(y_{i}|{\zeta}_{i},\sigma_{0}(1-\rho))}, \\
    K_{i}^{+}(\zeta_0,\zeta)&= \int p(y_{i}|{\zeta_{0,i}},\sigma_{0}) \log_{+} \frac{p(y_{i}|{\zeta_{0,i}},\sigma_{0})}{p(y_{i}|\zeta_{i},\sigma_{0}(1-\rho))} \diff y_{i},\\
    K_{i}^{-}(\zeta_0,\zeta)&= \int p(y_{i}|{\zeta_{0,i}},\sigma_{0}) \log_{-} \frac{p(y_{i}|{\zeta_{0,i}},\sigma_{0})}{p(y_{i}|\zeta_{i},\sigma_{0}(1-\rho))} \diff y_{i}.
\end{align*}
Then we get
\begin{align*}
&\mathbb{V}ar_{\zeta_0, \sigma_0}(W_i) = \mathbb{E}_{\zeta_0, \sigma_0}(W_i^2) - \{  K_{i}^{+}(\zeta_0,\zeta) \}^2\\
& \leq \mathbb{E}_{\zeta_0, \sigma_0}(W_i^2) - \{  K_{i}(\zeta_0,\zeta) \}^2\\
&\leq \mathbb{E}_{\zeta_0, \sigma_0}(W_i^2) + \int p(y_{i}|{\zeta_{0,i}},\sigma_{0}) \left( \log_{-} \frac{p(y_{i}|{\zeta_{0,i}},\sigma_{0})}{p(y_{i}|\zeta_{i},\sigma_{0}(1-\rho))} \right)^{2} \diff y_{i} - \{  K_{i}(\zeta_0,\zeta) \}^2\\
& = \int p(y_{i}|{\zeta_{0,i}},\sigma_{0}) \left( \log \frac{p(y_{i}|{\zeta_{0,i}},\sigma_{0})}{p(y_{i}|\zeta_{i},\sigma_{0}(1-\rho))} \right)^{2} \diff y_{i} - \{  K_{i}(\zeta_0,\zeta) \}^2\\
& = V_i(\zeta_0,\zeta).
\end{align*}
Hence, by condition (i) of (A1) for any $\rho < \tilde{\epsilon}_1$ and $\zeta \in B$
\begin{equation*}
    \sum_{i=1}^{\infty}\frac{\mathbb{V}ar_{\zeta_0, \sigma_0}(W_i)}{i^2} \leq \sum_{i=1}^{\infty}\frac{V_i(\zeta_0,\zeta)}{i^2} < \infty,
\end{equation*}
and by the Kolmogorov's strong law of large numbers for independent non-identically distributed random variables (e.g. \cite{Shiryaev1996}, Chapter 3),
\begin{equation*}
    \frac{1}{n}\sum_{i=1}^{n}(W_{i}-K_{i}^{+}(\zeta_0,\zeta)) \xrightarrow[\text{n}]{\quad} 0 \text{ a.s. } P_0.
\end{equation*}
As a result, for every $\zeta \in B$, with $P_0$ probability 1
\begin{align*}
    &\liminf_{n \rightarrow \infty} \bigg(\frac{1}{n}\sum_{i=1}^{n} \log \frac{p(y_{i}|\zeta_{i},\sigma_{0}(1-\rho))}{p(y_{i}|{\zeta_{0,i}},\sigma_{0})}  \bigg) \\
    &=-\liminf_{n \rightarrow \infty} \left(\frac{1}{n}\sum_{i=1}^{n} -\log \frac{p(y_{i}|\zeta_{i},\sigma_{0}(1-\rho))}{p(y_{i}|{\zeta_{0,i}},\sigma_{0})} \right) \\
    &=-\liminf_{n \rightarrow \infty} \left(\frac{1}{n}\sum_{i=1}^{n} \log \frac{p(y_{i}|{\zeta_{0,i}},\sigma_{0})}{p(y_{i}|\zeta_{i},\sigma_{0}(1-\rho))} \right) \\
    &\geq -\limsup_{n \rightarrow \infty} \left(\frac{1}{n}\sum_{i=1}^{n} \log_{+} \frac{p(y_{i}|{\zeta_{0,i}},\sigma_{0})}{p(y_{i}|\zeta_{i},\sigma_{0}(1-\rho))}\right)\\
    &=-\limsup_{n \rightarrow \infty}\left(\frac{1}{n}\sum_{i=1}^{n} K_{i}^{+}(\zeta_0,\zeta)\right)\\
    &\geq -\limsup_{n \rightarrow \infty}\left(\frac{1}{n}\sum_{i=1}^{n} K_{i}(S_0,S)+\frac{1}{n}\sum_{i=1}^{n} \sqrt{\frac{K_{i}(\zeta_0,\zeta)}{2}} \; \right) \\
    &\geq -\limsup_{n \rightarrow \infty}\left(\frac{1}{n}\sum_{i=1}^{n} K_{i}(\zeta_0,\zeta)+\sqrt{\frac{1}{n}\sum_{i=1}^{n} \frac{K_{i}(\zeta_0,\zeta)}{2}} \; \right).
\end{align*}
The fourth line follows from the almost sure convergence proved in the previous paragraph, and the second to last line follows from \cite{amewou-atisso2003}. We now make use of the condition (ii) of (A1). Let us consider $\beta > 0$ and select $\Delta$ so that $\Delta + \sqrt{\frac{\Delta}{2}} \leq \frac{\beta}{8}$ and also $C = B \cap \{\zeta: K_i(\zeta_0,\zeta) < \Delta\ \text{ for all  }i\}$. By (A1) there exists $\tilde{\epsilon}_1$ so that for all $0<\rho<\tilde{\epsilon}_1$ implies $\Pi(C|\thetab, \phib) > 0$. Therefore, for each $\zeta \in C$
\begin{align*}
    &\liminf_{n \rightarrow \infty} \left(\frac{1}{n}\sum_{i=1}^{n} log \frac{p(y_{i}|\zeta_{i},\sigma_{0}(1-\rho))}{p(y_{i}|{\zeta_{0,i}},\sigma_{0})} \right)    \\
     &\geq -\limsup_{n \rightarrow \infty}\left(\frac{1}{n}\sum_{i=1}^{n} K_{i}(\zeta_0,\zeta)+\sqrt{\frac{1}{n}\sum_{i=1}^{n} \frac{K_{i}(\zeta_0,\zeta)}{2}} \;\right)\\
     &\geq -(\Delta+\sqrt{\frac{\Delta}{2}}),
\end{align*}
since $\frac{1}{n}\sum_{i=1}^{n} K_{i}(\zeta_0,\zeta) <\Delta$ for all $\zeta \in C$. Finally, for any $\rho < \min \{\tilde{\epsilon}_1, \frac{1-e^{\frac{-\beta}{8}}}{1+e^{\frac{-\beta}{8}}}\}$

\small
\begin{align*}
      &\liminf_{n \rightarrow \infty}e^{\frac{2n\beta}{8}} \mathbf{I}_{3n}(y_1, \dots, y_n, \thetab, \phib, \hat{\sigma}_n) \\
      &\ge  \liminf_{n \rightarrow \infty}e^{\frac{2n\beta}{8}} {\left(\frac{1-\rho}{1+\rho}\right)}^{n} \int_{\mathcal{F}} \prod_{i = 1}^n\frac{p(y_i|\zeta_i,\sigma_0(1-\rho))}{p(y_i|\zeta_{0,i},\sigma_0)} \diff \Pi(\zetab|\thetab, \phib)\\
      &\ge  \liminf_{n \rightarrow \infty}e^{\frac{2n\beta}{8}} {\left(\frac{1-\rho}{1+\rho}\right)}^{n} \int_{C} \prod_{i = 1}^n\frac{p(y_i|\zeta_i,\sigma_0(1-\rho))}{p(y_i|\zeta_{0,i},\sigma_0)} \diff \Pi(\zetab|\thetab, \phib)\\
      &\ge \int_{C} \liminf_{n \rightarrow \infty}e^{\frac{2n\beta}{8}} {\left(\frac{1-\rho}{1+\rho}\right)}^{n}  \prod_{i = 1}^n\frac{p(y_i|\zeta_i,\sigma_0(1-\rho))}{p(y_i|\zeta_{0,i},\sigma_0)} \diff \Pi(\zetab|\thetab, \phib)\\
      & = \infty.
\end{align*}
\normalsize
Note that the actual bound on $\mathbf{I}_{3n}$ does not depend on $(\thetab, \phib)$. 
Taking $\epsilon < \min\{\tilde{\epsilon}_2, \frac{1 - e^{-c_1}}{1 + e^{-c_1}}\}$ concludes the proof.

\section{Proof of Lemma \ref{lemma:consistency:prior}}\label{sec:Appendix:lemma}
Theorem 5 of \cite{ghosal2006} implies that there exist positive constants $C, d_1, \dots, d_p$ so that for $i = 1, \dots, p$
\begin{align*}
    P\left( \sup_{ \tb \in [0,1]^p}|\zeta(\tb)|>M_{n}\bigg | \zb, \thetab, \phib, \right) &\leq Ce^{-d_0 \frac{M_{n}^2}{\rho_{0}^{2}(\thetab, \phib)}}, \\
    P\left( \sup_{ \tb \in [0,1]^p}\Big|\frac{\partial}{\partial t_{i}}\zeta(\tb)\Big|>M_{n}| \zb, \thetab, \phib, \right) &\leq Ce^{-d_i \frac{M_{n}^2}{\rho_{i}^{2}(\thetab, \phib)}}.
\end{align*}

The continuity of $\rho^2_i(\thetab, \phib)$, for $i = 0,\cdots, p$, on a compact set $\Upsilon$ implies that they are uniformly bounded. Therefore, there exist universal constants $(c_{0,1},c_{0,2}), \cdots$, $(c_{p,1}, c_{p,2})$ such that for $i = 0, \cdots, p$,
\begin{equation*}
    0 < c_{i,1} \leq \sup_{(\thetab, \phib) \in \Upsilon} |\rho_i^2(\thetab, \phib)| \leq c_{i,2}.
\end{equation*}
Hence, for $i = 0, \cdots, p$,
\begin{align*}
   \sup_{(\thetab, \phib) \in \Upsilon} P\left( \sup_{ \tb \in [0,1]^p}|\zeta(\tb)|>M_{n}\bigg | \zb, \thetab, \phib, \right) &\leq Ce^{-d_0 \frac{M_{n}^2}{c_{0,1}}}, \\
   \sup_{(\thetab, \phib) \in \Upsilon} P\left( \sup_{ \tb \in [0,1]^p}\Big|\frac{\partial}{\partial t_{i}}\zeta(\tb)\Big|>M_{n}| \zb, \thetab, \phib, \right) &\leq Ce^{-d_i \frac{M_{n}^2}{c_{i,1}}}.
\end{align*}

\section{Proof of Lemma \ref{thrm:tests}}\label{sec:Appendix:tests}
We shall first define some notation. Let $0 <r <\frac{\nu}{2}$ and $t = \frac{r}{4}$. Let $N_t = N(t, \mathcal{F}_n, \parallel \cdot \parallel _{\infty})$ be the covering number of $\mathcal{F}_n$. In Theorem 2.7.1, \cite{van1996weak} show that there exist a constant $K$ so that $\log N_t \leq \frac{KM_n}{t^p}$ and therefore $N_t = \mathcal{O}(M_n)$, where $M_n = \mathcal{O}(n^\alpha)$ for $\alpha \in (\frac{1}{2},1)$ according to the definition of the sieves. Let us consider $\tau \in (\frac{\alpha}{2}, \frac{1}{2})$ and define $c_n = n^{\tau}$ so that $\log(N_t) = 
o(c_n^2)$. Moreover, let $\zeta^1, \dots, \zeta^{N_t} \in \mathcal{F}_n$ be finitely many elements of the sieve so that for every $\zeta \in \mathcal{F}_n$ there is $i \in \{1, \dots, N_t\}$ satisfying  $\parallel\zeta - \zeta^i \parallel_\infty < t$. This implies that if $\zeta \in \mathcal{F}_n$ such that $\int|\zeta(\tb) - \zeta_0(\tb)| \diff Q_n(\tb) > \nu$, then $\int|\zeta^i(\tb) - \zeta_0(\tb)| \diff Q_n(\tb) >  \frac{\nu}{2}$.

The next step in the proof is to construct a test for each $\zeta^i$ with the resulting functions $\Phi_n$ defined as a combination of the individual tests and showing that the probabilities of type I and type II errors satisfies the properties of the lemma. Let us recall that $\zeta_j = \zeta(\tb_j)$ and $\zeta_{0,j} = \zeta_0 (\tb_j)$. For an arbitrary $\zeta \in \mathcal{F}_n$ such that $\parallel\zeta - \zeta^i \parallel_\infty < t$, let us define $\zeta_{1,j} = \zeta^i(\tb_j)$ and $b_j = 1$ if $\zeta_{1,j} > \zeta_{0,j}$ and $-1$ otherwise. For any $\nu > 0$, let $\Psi_n[\zeta, \nu]$ be the indicator of set $A$ defined as follows
\begin{equation*}
    A = \left\{  \sum_{j=1}^{n}b_{j}\left(\frac{y_{j}-\zeta_{0,j}}{\sigma_0}\right) > 2c_{n}\sqrt{n} \right\}.
\end{equation*}
The test functions $\Phi_n$ are then
\begin{equation*}
    \Phi_{n}= \max_{1\leq j \leq N_{t}}\Psi_{n}[\zeta^{j},\frac{\nu}{2}].
\end{equation*}

\paragraph{Type I error)} The Mill's ratio implies
\begin{align*}
    \mathbb{E}_{\zeta_{0},\sigma_0}(\Psi_{n}) &=P_{0}\left[\sum_{j=1}^{n}b_{j}\left(\frac{y_{j}-\zeta_{0,j}}{\sigma_0}\right) > 2c_{n}\sqrt{n}\right] \\
     &=1-\Phi(2c_n)\\
     &\leq \frac{1}{2c_n\sqrt{2\pi}} e^{-2c_n^2}\\
     & \leq e^{-2c_n^2}.
\end{align*}
The function $\Phi(\cdot)$ is the CDF of the standard normal distribution. Consequently, we have
\begin{align*}
    \mathbb{E}_{\zeta_{0},\sigma_0}(\Phi_{n}) &\leq \sum_{j =1}^{N_t} \mathbb{E}_{\zeta_{0},\sigma_0}(\Psi_{n}[\zeta^{j},\frac{\nu}{2}])\\
    &\leq N_t e^{-2c_n^2} = e^{\log(N_t)-2c_n^2}\\
    &\leq e^{-c_n^2},
\end{align*}
and
\begin{equation*}
    \sum_{n=1}^{\infty}\mathbb{E}_{\zeta_{0},\sigma_0}\Phi_{n} < \infty.
\end{equation*}
\paragraph{Type II error)}
It is sufficient to find $i$ for which the probability of type II error of $\Psi_n[\zeta^i, \frac{\nu}{2}]$, given an arbitrary $\zeta$ in $W^C_{\nu, n} \cap \mathcal{F}_n$, is sufficiently small. This is because the probability of type II error for the composite test $\Phi_n$ is no larger than the smallest of $\Psi_n[\zeta^i, \frac{\nu}{2}]$. Note that here we assume $\int|\zeta(\tb) - \zeta_0(\tb)| \diff Q_n(\tb) > \nu$, and then $\int|\zeta^i(\tb) - \zeta_0(\tb)| \diff Q_n(\tb) >  \frac{\nu}{2}$. For every $r < \frac{\nu}{2}$, \cite{Choi2007} show that
\begin{equation*}
\sum_{j=1}^n |\zeta_{1,j} - \zeta_{0,j}| > rn.
\end{equation*}

Let $n$ be large enough so that $4\sigma_0 c_n < r\sqrt{n}$, then for any $0<\epsilon<1$

\small
\begin{align*}
    &\mathbb{E}_{\zeta, \sigma_0 (1 + \epsilon)}(1 - \Psi_n[\zeta^i, \frac{\nu}{2}]) \\
    &= P_{\zeta, \sigma_0 (1 + \epsilon)} \Bigg[\sum_{j=1}^{n}b_{j}\left(\frac{y_{j}-\zeta_{0,j}}{\sigma_0}\right) \leq 2c_{n}\sqrt{n}\Bigg] \\
    &=P_{\zeta, \sigma_0 (1 + \epsilon)} \Bigg[\frac{1}{\sqrt{n}}\sum_{j=1}^{n}b_{j}\left(\frac{y_{j}-\zeta_{j}}{\sigma_0}\right) + \frac{1}{\sqrt{n}}\sum_{j=1}^{n}b_{j}\left(\frac{\zeta_{j}-\zeta_{1,j}}{\sigma_0}\right)+ \frac{1}{\sqrt{n}}\sum_{j=1}^{n}\left|\frac{\zeta_{1,j}-\zeta_{0,j}}{\sigma_0}\right| \leq 2 c_n \Bigg]\\
    &\leq P_{\zeta, \sigma_0 (1 + \epsilon)} \Bigg[ \frac{1}{\sqrt{n}}\sum_{j=1}^{n}b_{j}\left(\frac{y_{j}-\zeta_{j}}{\sigma_0}\right) \leq \frac{r \sqrt{n}}{4 \sigma_0} - \frac{r \sqrt{n}}{\sigma_0}  + 2c_n\Bigg]\\
    &\leq  P_{\zeta, \sigma_0 (1 + \epsilon)} \Bigg[ \frac{1}{\sqrt{n}}\sum_{j=1}^{n}b_{j}\left(\frac{y_{j}-\zeta_{j}}{\sigma_0 (1+\epsilon)}\right) \leq - \frac{r \sqrt{n}}{4 \sigma_0 (1+\epsilon)} \Bigg]\\
    &=\Phi \left( - \frac{r \sqrt{n}}{4 \sigma_0 (1+\epsilon)} \right)\\
    &\leq \frac{4\sigma_0(1+\epsilon)}{r \sqrt{2 \pi n}}e^{-\frac{nr^2}{32 \sigma_0^2 (1+\epsilon)^2}}.
\end{align*}
\normalsize

To establish the part (ii) of the lemma, we need to show that there exists $0<\tilde{\epsilon}<1$ so that for any $\epsilon<\tilde{\epsilon}$
\begin{equation}\label{eqn:EB:thrm2:final:step}
    \frac{r^2}{32 \sigma_0^2 (1+\epsilon)^2} + \log\left(\frac{1-\epsilon}{1+\epsilon}\right) > 0.
\end{equation}
Take $\kappa = \frac{r^2}{32 \sigma_0^2}$ and define $b(\epsilon)$ to be the left hand side of \eqref{eqn:EB:thrm2:final:step},
\begin{equation*}
    b(\epsilon) = \kappa\left(\frac{1}{(1+\epsilon)^2}  + \frac{1}{\kappa} \log\left(\frac{1-\epsilon}{1+\epsilon}\right) \right).
\end{equation*}
The function $b(\epsilon)$ is clearly continuous at $\epsilon = 0$. Hence, for each $\kappa > 0$, there exists $\tilde{\epsilon}$ such that for all $0<\epsilon<\tilde{\epsilon}$, $b(\epsilon) > 0$.

\section{Proof of Theorem \ref{thrm:Sigma}}\label{sec:Appendix:Sigma}
First, we show that $\hat{\sigma}_n^2$ is asymptotically unbiased. Note that
\begin{align*}
   \mathbb{E}[(y_{i+1} - y_i)^2] &=  [\zeta_0(\tb_{i + 1}) - \zeta_0(\tb_{i})]^2 + \sigma_0^2\mathbb{E}[(\epsilon_{i+1} - \epsilon_i)^2] \\
   &= [\zeta_0(\tb_{i + 1}) - \zeta_0(\tb_{i})]^2 + 2\sigma_0^2,
\end{align*}
because $\epsilon_i \stackrel{i.i.d.}{\sim} \cN(0,1)$. Consequently
\begin{equation}\label{eqn:ConsistenSigmaExp}
   \mathbb{E}(\hat{\sigma}_n^2) =\frac{\sum_{i = 1}^{n-1}[\zeta_0(\tb_{i + 1}) - \zeta_0(\tb_{i})]^2}{2(n - 1)} + \sigma_0^2.
\end{equation}
Since $\zeta_0$ is continuously differentiable on the compact and convex set $\Omegab$, it is also (globally) Lipschitz on $\Omegab$ (e.g. \cite{Schaeffer2016}, Corollary 3.2.4), and there exist a real constant $K$ so that
\begin{align*}
    |\zeta_0(\tb_{i + 1}) - \zeta_0(\tb_{i})| \le K \sum_{j = 1}^ p |t_{i+1,j} - t_{i,j}|.
\end{align*}
Therefore, due to the design assumption \eqref{eqn:DesignReq}
\begin{equation}\label{eqn:ConsistenSigmaExpLem}
\begin{split}
    0 &\le \frac{\sum_{i = 1}^{n-1}[\zeta_0(\tb_{i + 1}) - \zeta_0(\tb_{i})]^2}{2(n - 1)} \\
    &\le \frac{K^2p^2}{2} \bigg[\sup_{i \in \{1, \dots, n\}, j \in \{1, \dots, p\}} |t_{i+1, j} - t_{i,j}|\bigg]^2 \xrightarrow[\text{n}]{\quad \quad} 0,
    \end{split}
\end{equation}
and the combination of \eqref{eqn:ConsistenSigmaExp} with \eqref{eqn:ConsistenSigmaExpLem} implies
\begin{equation}\label{eqn:SigmaUnbiased}
     \mathbb{E}(\hat{\sigma}_n^2) \xrightarrow[\text{n}]{\quad \quad} \sigma_0^2.
\end{equation}

To show the almost sure convergence of $\hat{\sigma}_n^2$, let us now denote $x_i = (y_{i + 1} - y_i)^2$ and rewrite the estimator $\hat{\sigma}_n^2$ as a sum of two estimators, each consisting of a sum of independent variables:
\begin{align*}
    \hat{\sigma}_n^2 = \frac{\frac{1}{2} \sum_{i = 1}^{\frac{n -1 }{2}} x_{2i} }{2\big(\frac{n -1 }{2}\big)} +  \frac{\frac{1}{2} \sum_{j = 1}^{\frac{n -1 }{2}} x_{2j-1} }{2\big(\frac{n -1 }{2}\big)} = \hat{\sigma}_{n,e}^2 + \hat{\sigma}_{n,o}^2.
\end{align*}
Without loss of generality, we assumed that $n$ is an odd integer. Lastly note that $\mathbb{V}ar(x_i) \le C < \infty$ uniformly in $i$. This is because the differences $\zeta_0(\tb_{i+1}) - \zeta_0(\tb_{i})$ are uniformly bounded on the compact set $\Omegab$ due to the continuity of $\zeta_0$. Additionally, $y_{i+1} - y_i$ are normal and have bounded moments. We can now apply the Kolmogorov's strong law of large numbers for independent non-identically distributed random variables (e.g. \cite{Shiryaev1996}, Chapter 3),
\begin{align*}
    &\hat{\sigma}_{n,e}^2 \xrightarrow[\text{n}]{\quad \quad} \frac{1}{2}\sigma_0^2 \quad \text{a.s. } P_0 \\ 
    &\hat{\sigma}_{n,0}^2 \xrightarrow[\text{n}]{\quad \quad} \frac{1}{2}\sigma_0^2 \quad \text{a.s. } P_0
\end{align*}
and as a result
\begin{equation*}
    \hat{\sigma}_n^2 = \hat{\sigma}_{n,e}^2 + \hat{\sigma}_{n,o}^2 \xrightarrow[\text{n}]{\quad \quad} \sigma_0^2 \quad \text{a.s. } P_0.
\end{equation*}

\section{The LDM calibration}\label{sec:app:LDM}
The analysis of the LDM follows our previous study in \cite{kejzlar2020variational}. Here we provide a concise discussion regarding the choices of prior distributions for and the GP's specification.
\paragraph{GP specifications.} For the computer model $E_{\textrm{B}}(Z,N)$, we consider the GP prior with the mean zero and the covariance function

\small
\begin{align*}
	\eta_E \cdot \text{exp}&(-\frac{(Z - Z' )^2}{2\nu^2_Z} -\frac{(N - N')^2}{2\nu^2_N} -\frac{(\theta_{\textrm{vol}} - \theta_{\textrm{vol}}')^2}{2\nu^2_1} \\
	&- \frac{(\theta_{\textrm{surf}} - \theta_{\textrm{surf}}' )^2}{2\nu^2_2} -\frac{(\theta_{\textrm{sym}} - \theta_{\textrm{sym}}' )^2}{2\nu^2_3} -\frac{(\theta_{\textrm{C}} - \theta_{\textrm{C}}' )^2}{2\nu^2_4}).
\end{align*}
\normalsize
We also assume the GP prior for the systematic discrepancy $\delta(Z,N)$ with mean zero and covariance function 
\begin{equation*}
	\eta_\delta \cdot \text{exp}(-\frac{(Z - Z' )^2}{2l^2_Z} -\frac{(N - N' )^2}{2l^2_N}).
\end{equation*}

\paragraph{Prior distributions} The prior distributions for the calibration parameters $\thetab$ are chosen to be wide enough to cover the space of all their reasonable values:
	\begin{align*}\label{eqn:priors_theta}
	\theta_{\textrm{vol}} &\sim \cN(15.42, 0.203), \\
	\theta_{\textrm{surf}} &\sim \cN(16.91, 0.645), \\
	\theta_{\textrm{sym}} &\sim \cN(22.47, 0.525), \\
	\theta_{\textrm{C}} &\sim \cN(0.69, 0.015).
\end{align*}
The prior distributions for the hyperparameters $\phib$ were selected as $Gamma(\alpha, \beta)$ with the shape parameter $\alpha$ and scale parameter $\beta$. They are chosen to be weakly informative so that they correspond to the scale of these parameters given by the literature on nuclear mass models \citep{Weizsacker1935,Bethe36,Myers1966, Fayans1998, Kirson2008, McDonnellPRL15, UNEDF0, UNEDF1, UNEDF2, Benzaid2020, Kejzlar2020}. In particular,
\begin{align*}
	\sigma &\sim Gamma(2,1),\\
	\eta_\delta &\sim Gamma(10,1),\\
	l_Z &\sim Gamma(10,1), \\
	l_N &\sim Gamma(10,1), \\
	\nu_Z &\sim Gamma(10,1), \\
	\nu_N &\sim Gamma(10,1), \\
	\nu_i &\sim Gamma(10,1), \hspace{1cm} i = 1,2,3,4.
\end{align*}

Since the majority of the masses in the training dataset are larger than 1000 MeV. We consider the following prior for $\eta_f$ to reflect this notion
\begin{equation*}
	\eta_f \sim Gamma(110,10).
\end{equation*}

\section{Numerical study of the conditional covariance $k_\zeta$}\label{sec:Appendix:kernel}

Here we present the numerical investigation of our conjecture about the asymptotic behavior of the conditional covariance $k_{\zeta}$. We show that with increasing number of model evaluations $s$ (assuming some space filling design) the covariance function $k_{\delta}$ quickly dominates which strongly points out to similar asymptotic behavior of $k_{\zeta}$ and $k_{\delta}$ with respect to $s$. Our rational is that by informing the prior distribution for $\zeta$ with more model evaluations, we effectively reduce the uncertainty about the computer model.

\subsection{Study design}
We consider a simple scenario with the joint space of model and calibration inputs over $[0,1]^2$. The input pairs $(\tilde{t}_j, \tilde{\theta}_j)$ were generated using the space filling Latin hypercube design. The true value of calibration parameter was chosen to vary between $\theta = \{0.3, 0.5, 0.8\}$.

\subsection{Results}
\begin{figure*}
	\begin{centering}
		\includegraphics[width=1\textwidth]{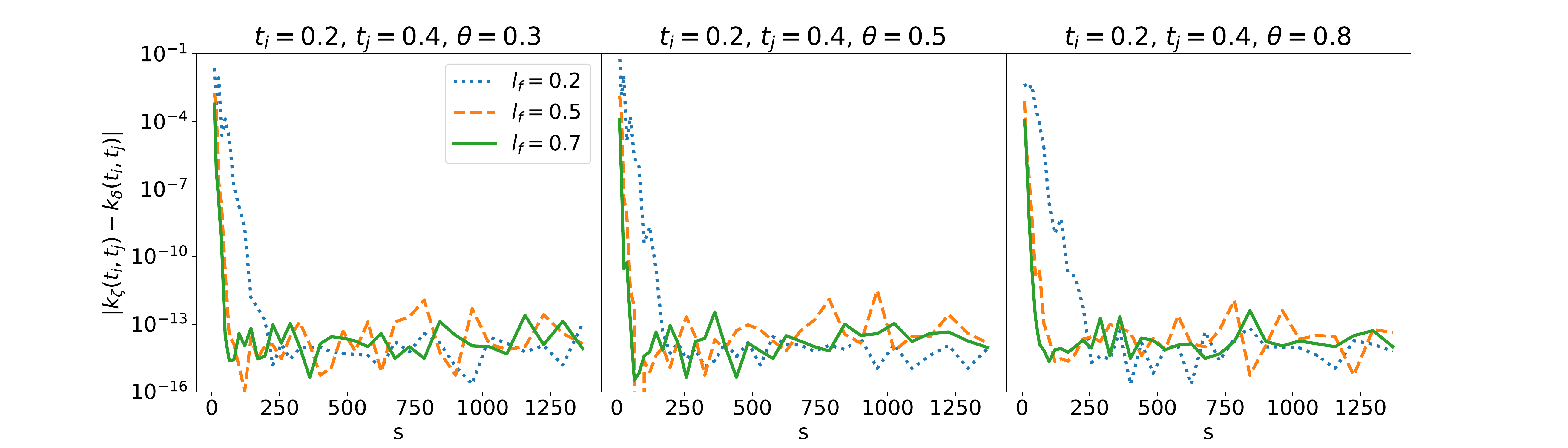}
		\caption{The absolute difference between the conditional kernel $k_{\zeta}$ and $k_{\delta}$ for the model inputs $t_i = 0.2$ and $t_j = 0.4$ and the value of true calibration parameter $\theta = \{0.3, 0.5, 0.8\}$. This is the squared exponential kernel case.}
		\label{fig:Num:1}
	\end{centering}
\end{figure*}

\begin{figure*}
	\begin{centering}
		\includegraphics[width=1\textwidth]{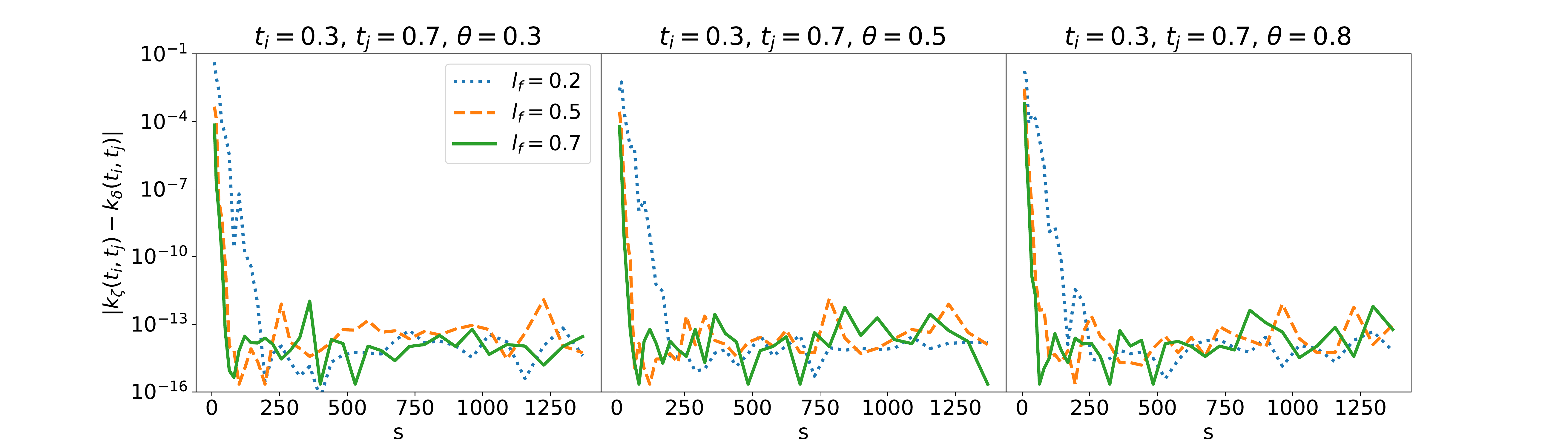}
		\caption{The absolute difference between the conditional kernel $k_{\zeta}$ and $k_{\delta}$ for the model inputs $t_i = 0.3$ and $t_j = 0.7$ and the value of true calibration parameter $\theta = \{0.3, 0.5, 0.8\}$. This is the squared exponential kernel case.}
		\label{fig:Num:2}
	\end{centering}
\end{figure*}

Figure \ref{fig:Num:1} and  Figure \ref{fig:Num:2} show the values of $| k_{\zeta}(t_i,t_j) -  k_{\delta}(t_i,t_j)|$ as a function of model runs in the case of squared exponential covariance kernels for both $k_f$ and $k_{\delta}$. The hyperparameter values were fixed to $\eta_f = \eta_{\delta} = 1$ with varying values for the length scales so that $l_f = l_{\delta}$. 

\begin{figure*}
	\begin{centering}
		\includegraphics[width=1\textwidth]{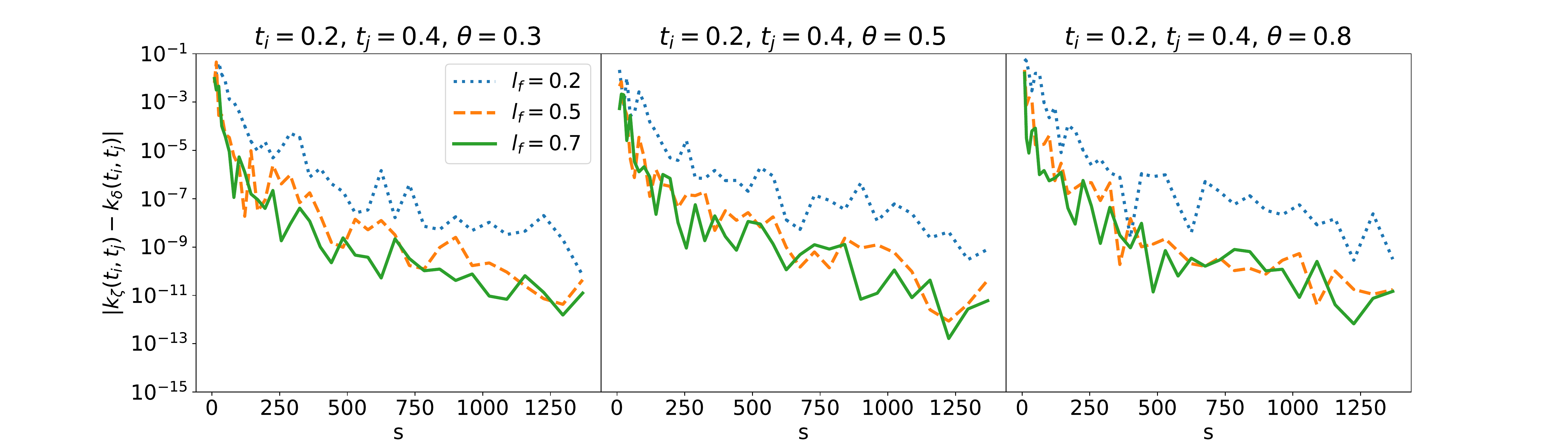}
		\caption{The absolute difference between the conditional kernel $k_{\zeta}$ and $k_{\delta}$ for the model inputs $t_i = 0.2$ and $t_j = 0.4$ and the value of true calibration parameter $\theta = \{0.3, 0.5, 0.8\}$. This is the tensor-product Mat\'ern case.}
		\label{fig:Num:3}
	\end{centering}
\end{figure*}

\begin{figure*}
	\begin{centering}
		\includegraphics[width=1\textwidth]{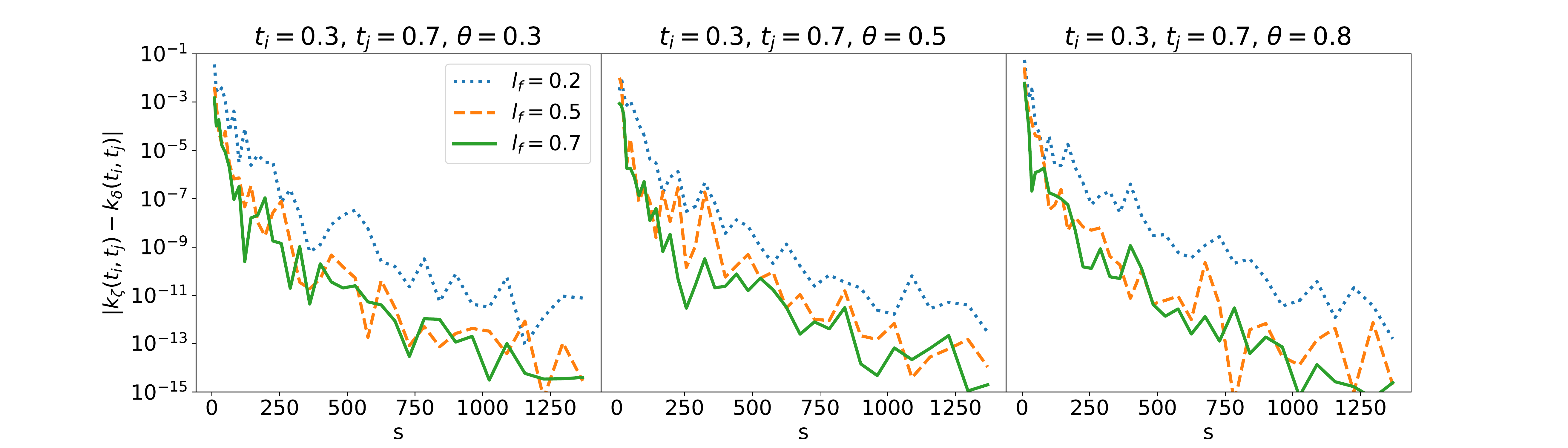}
		\caption{The absolute difference between the conditional kernel $k_{\zeta}$ and $k_{\delta}$ for the model inputs $t_i = 0.3$ and $t_j = 0.7$ and the value of true calibration parameter $\theta = \{0.3, 0.5, 0.8\}$. This is the tensor-product Mat\'ern case.}
		\label{fig:Num:4}
	\end{centering}
\end{figure*}

Analogically, Figures \ref{fig:Num:3} and \ref{fig:Num:4} correspond to the case of tensor-product Mat\'ern kernels with the standard choice of smoothness parameter $\lambda = 2.5$ which is sufficient according to the theory discussed in Section \ref{sec:Consistency:Posterior:assumptions}. The remaining hyperparameters are same as in the case of squared exponential kernel.

We can see that both choices of kernel functions for $k_f$ and $k_{\delta}$ exhibit the hypothesized dominance of $k_{\delta}$ with the increasing number of model runs. This happens irrespective of the choice of kernel function, model inputs $t$, calibration parameters $\theta$, and the length scales. Particularly in the case of squared exponential kernel, the absolute difference between $k_{\zeta}$ and $k_{\delta}$ quickly decreases and reaches the limits of numerical stability. On the other hand, the rate of convergence is considerably slower for the Mat\'ern kernel which is likely related to the limited smoothness of the kernel.

\newpage
\section{Additional results for simulation study: Transverse harmonic wave}\label{sec:Appendix:Sim1}

\begin{figure*}[h!]
	\begin{centering}
		\includegraphics[width=1\textwidth]{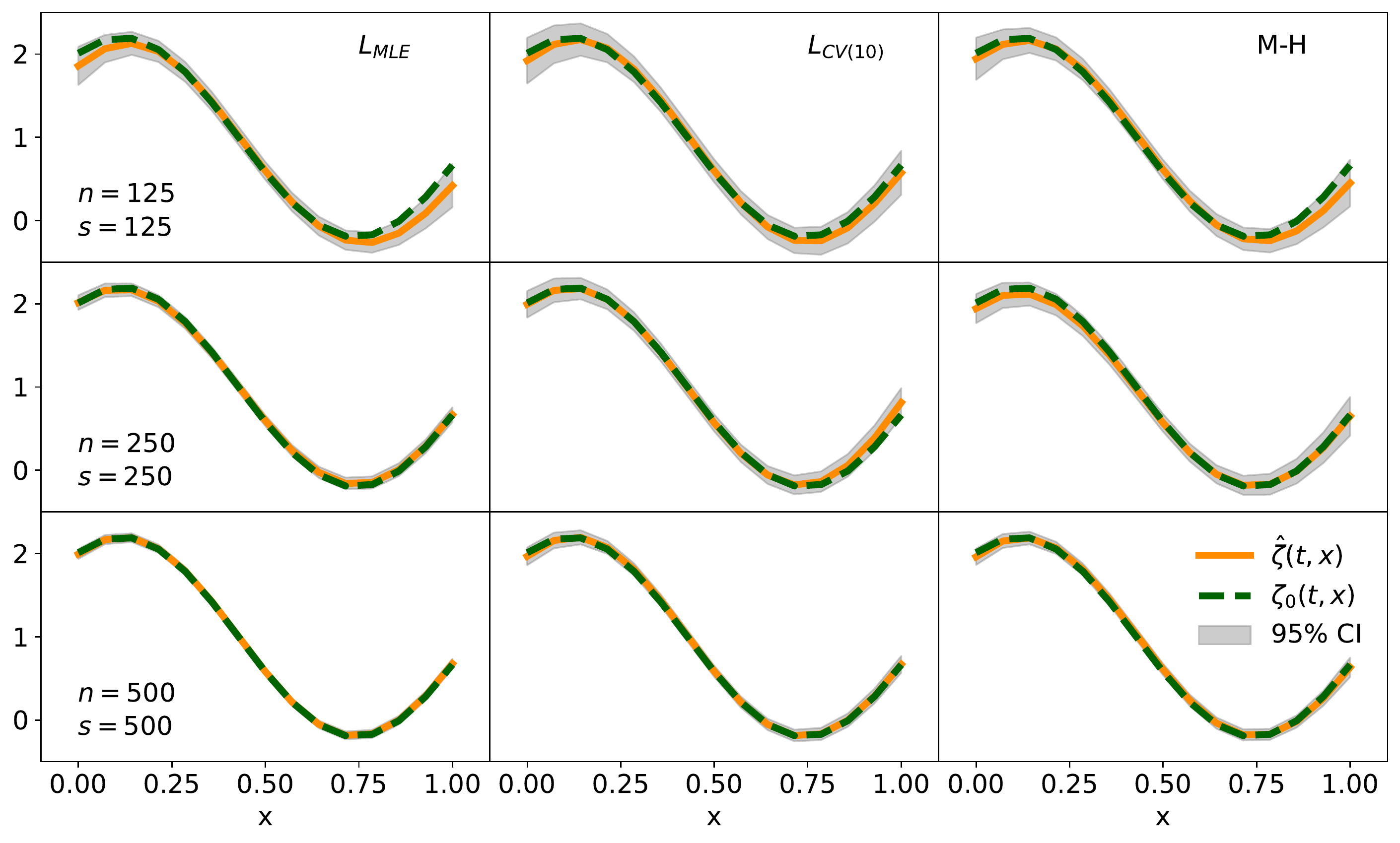}
		\caption{Comparison of the convergence to the true physical process $\zeta_0(t,x)$ under the empirical Bayes approach and the fully Bayesian implementation given by the Metropolis-Hastings algorithm. The dashed line represents the true process $\zeta_0$, and the solid line corresponds to the mean of posterior predictive distributions under respective method. The curves with $95 \%$ credible intervals (shaded area) are plotted at $t=0.00$.}
		\label{fig:Sim1:2}
	\end{centering}
\end{figure*}

\begin{figure*}
	\begin{centering}
		\includegraphics[width=1\textwidth]{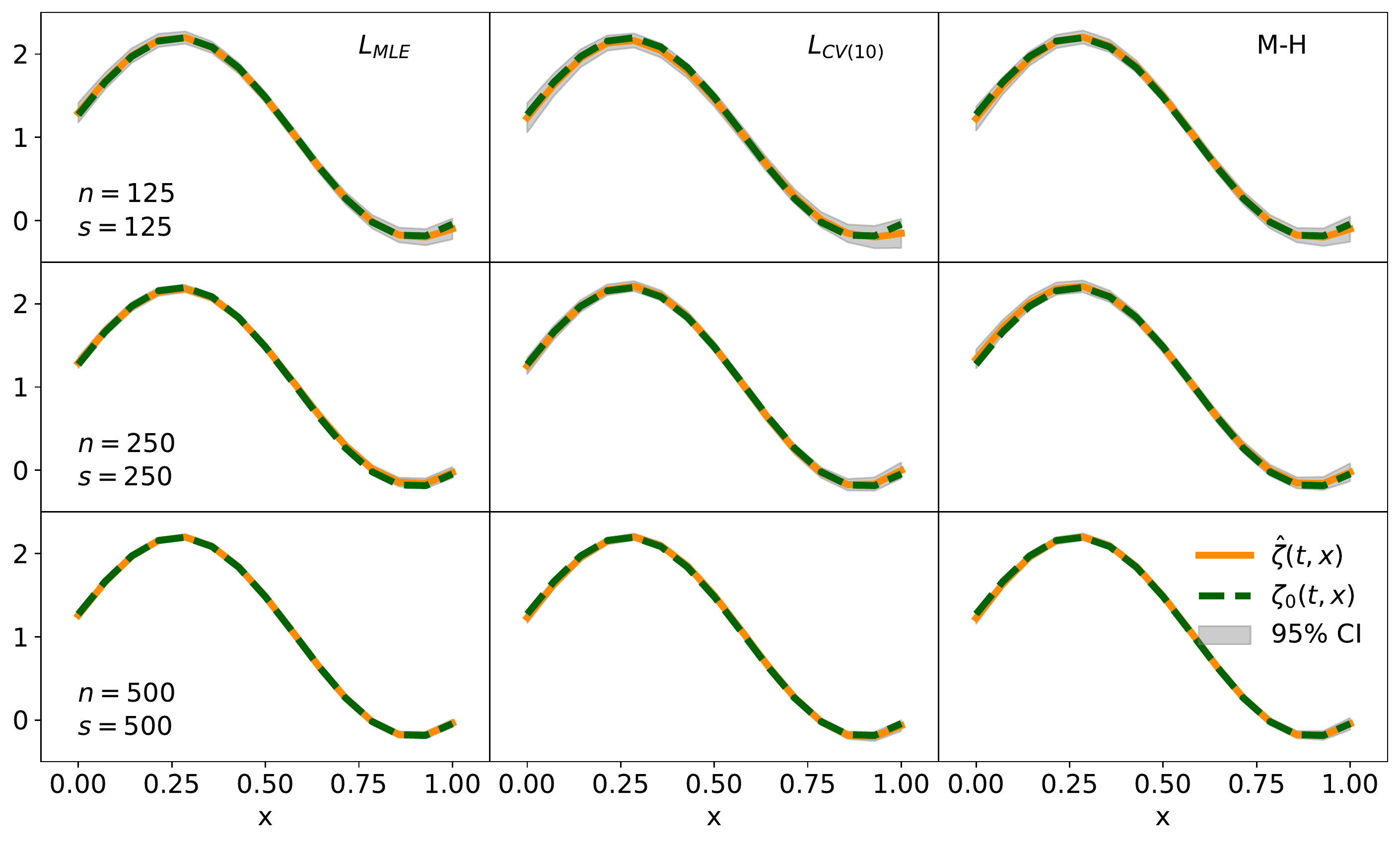}
		\caption{Comparison of the convergence to the true physical process $\zeta_0(t,x)$ under the empirical Bayes approach and the fully Bayesian implementation given by the Metropolis-Hastings algorithm. The dashed line represents the true process $\zeta_0$, and the solid line corresponds to the mean of posterior predictive distributions under respective method. The curves with $95 \%$ credible intervals (shaded area) are plotted at $t=0.43$.}
		\label{fig:Sim1:3}
	\end{centering}
\end{figure*}

\begin{figure*}
	\begin{centering}
		\includegraphics[width=1\textwidth]{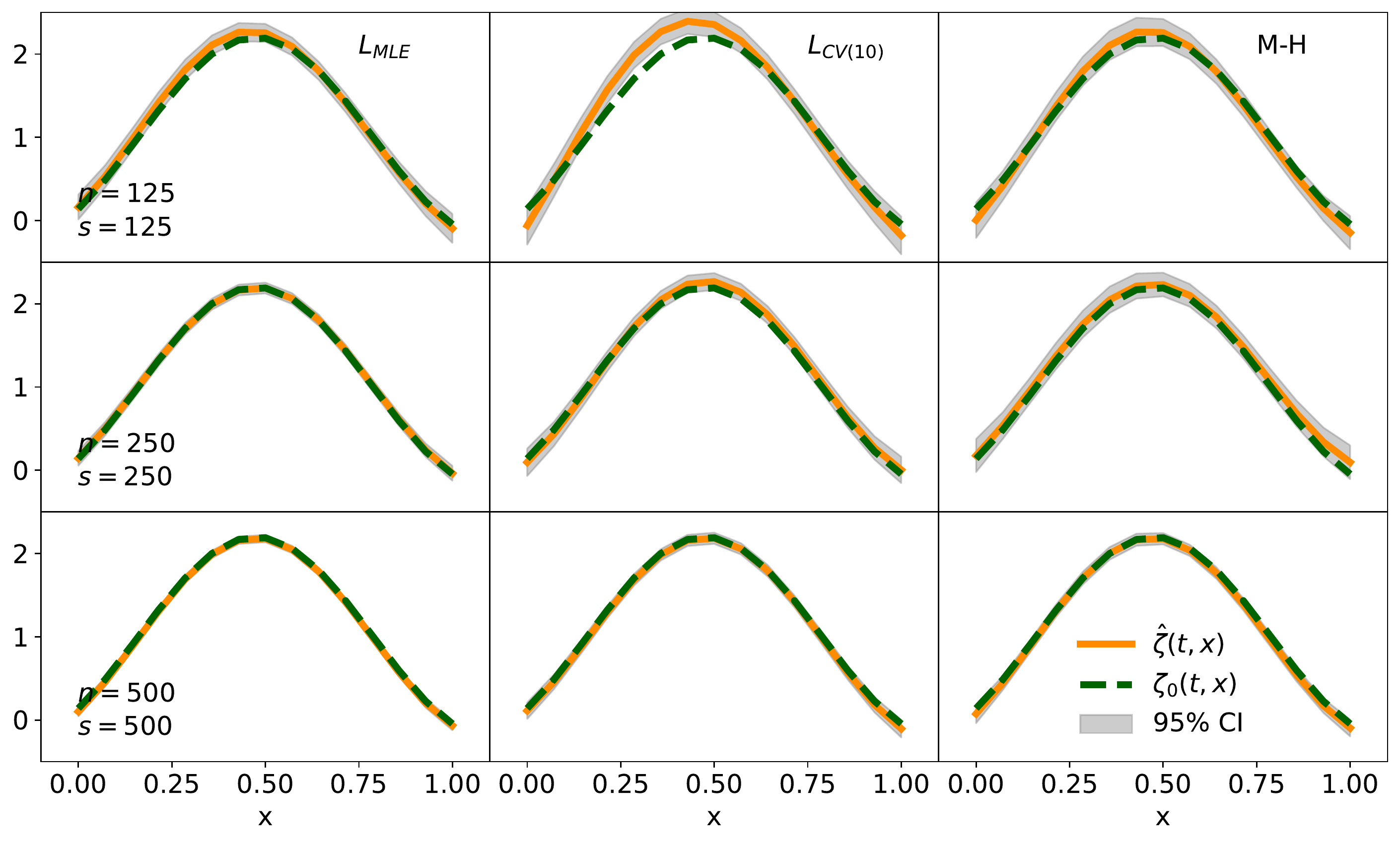}
		\caption{Comparison of the convergence to the true physical process $\zeta_0(t,x)$ under the empirical Bayes approach and the fully Bayesian implementation given by the Metropolis-Hastings algorithm. The dashed line represents the true process $\zeta_0$, and the solid line corresponds to the mean of posterior predictive distributions under respective method. The curves with $95 \%$ credible intervals (shaded area) are plotted at $t=0.71$.}
		\label{fig:Sim1:4}
	\end{centering}
\end{figure*}

\begin{figure*}
	\begin{centering}
		\includegraphics[width=1\textwidth]{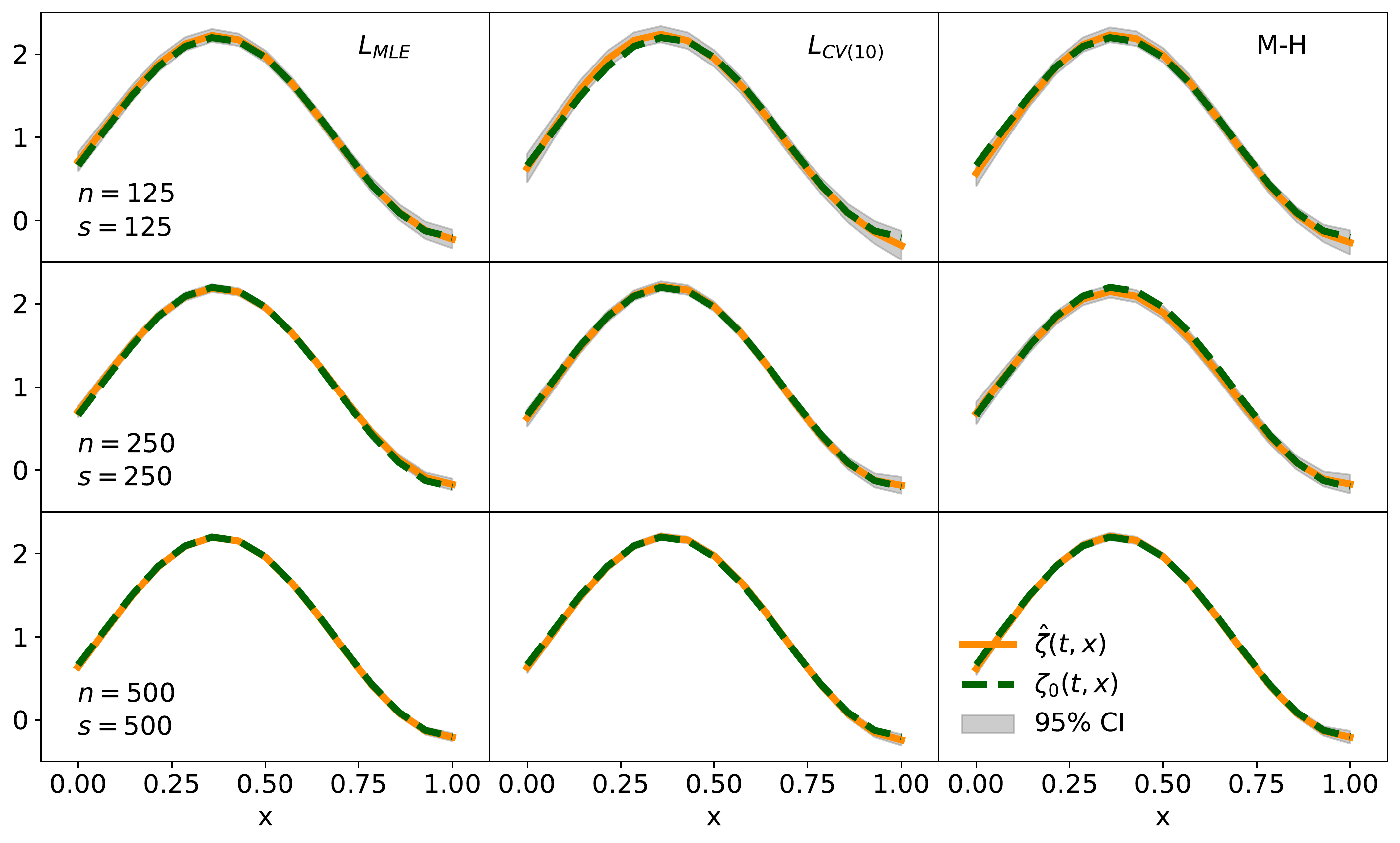}
		\caption{Comparison of the convergence to the true physical process $\zeta_0(t,x)$ under the empirical Bayes approach and the fully Bayesian implementation given by the Metropolis-Hastings algorithm. The dashed line represents the true process $\zeta_0$, and the solid line corresponds to the mean of posterior predictive distributions under respective method. The curves with $95 \%$ credible intervals (shaded area) are plotted at $t=1.00$.}
		\label{fig:Sim1:5}
	\end{centering}
\end{figure*}

\begin{figure*}
	\begin{centering}
		\includegraphics[width=1\textwidth]{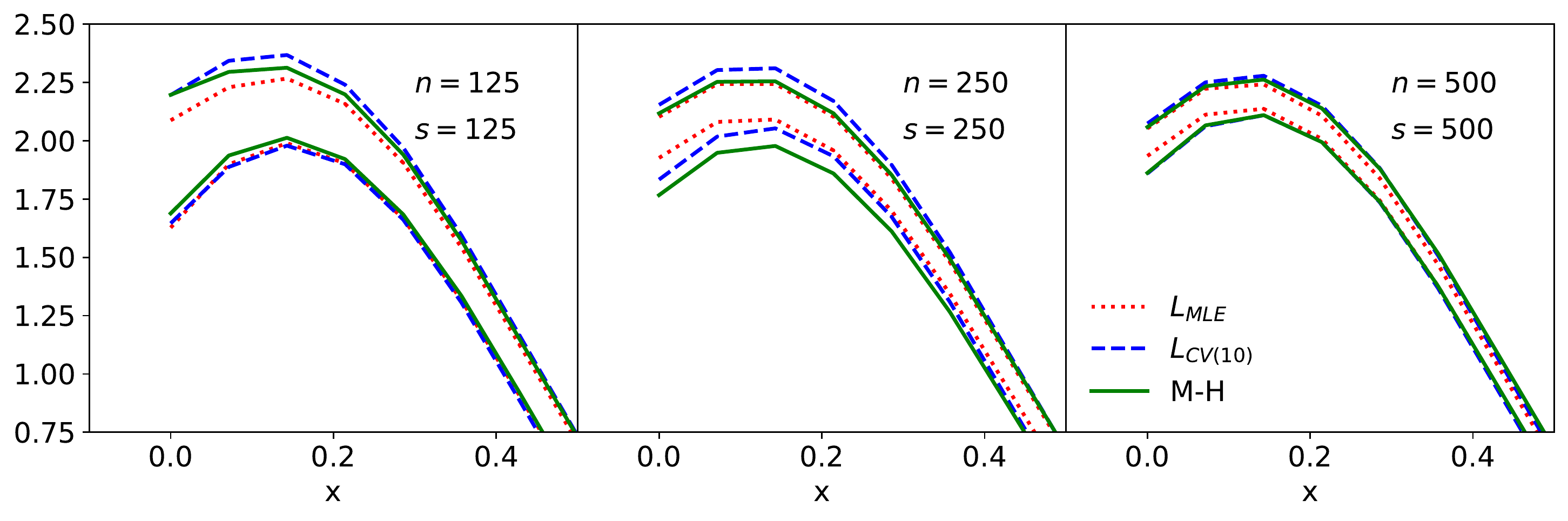}
		\caption{Details of $95 \%$ credible bands of posterior predictive distributions under the empirical Bayes approach and the fully Bayesian approach of Metropolis-Hastings algorithm. These were plotted at $t=0.00$.}
		\label{fig:Sim1:2:CB}
	\end{centering}
\end{figure*}

\begin{figure*}
	\begin{centering}
		\includegraphics[width=1\textwidth]{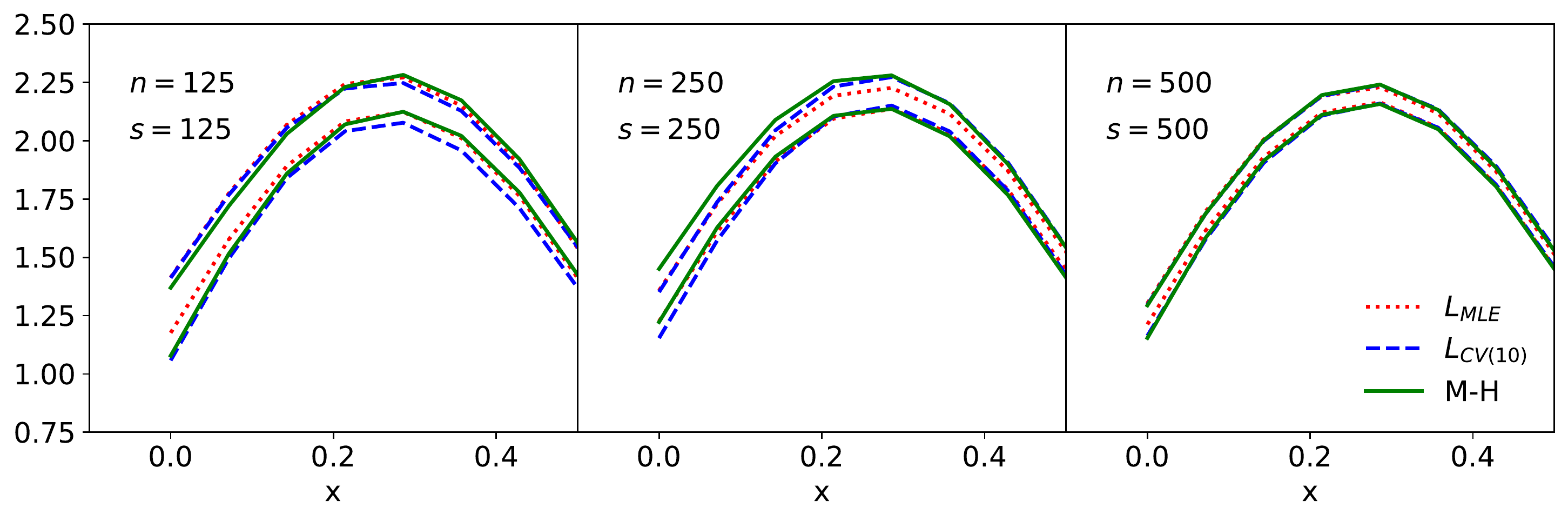}
		\caption{Details of $95 \%$ credible bands of posterior predictive distributions under the empirical Bayes approach and the fully Bayesian approach of Metropolis-Hastings algorithm. These were plotted at $t=0.43$.}
		\label{fig:Sim1:3:CB}
	\end{centering}
\end{figure*}

\begin{figure*}
	\begin{centering}
		\includegraphics[width=1\textwidth]{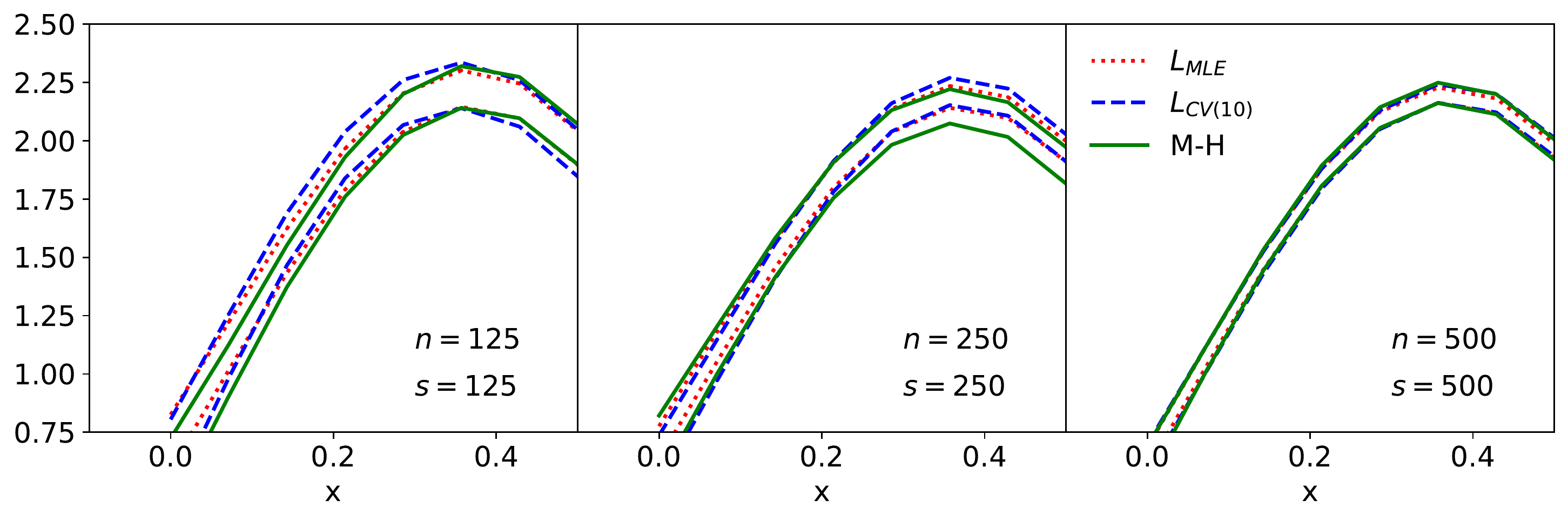}
		\caption{Details of $95 \%$ credible bands of posterior predictive distributions under the empirical Bayes approach and the fully Bayesian approach of Metropolis-Hastings algorithm. These were plotted at $t=0.71$.}
		\label{fig:Sim1:4:CB}
	\end{centering}
\end{figure*}

\begin{figure*}
	\begin{centering}
		\includegraphics[width=1\textwidth]{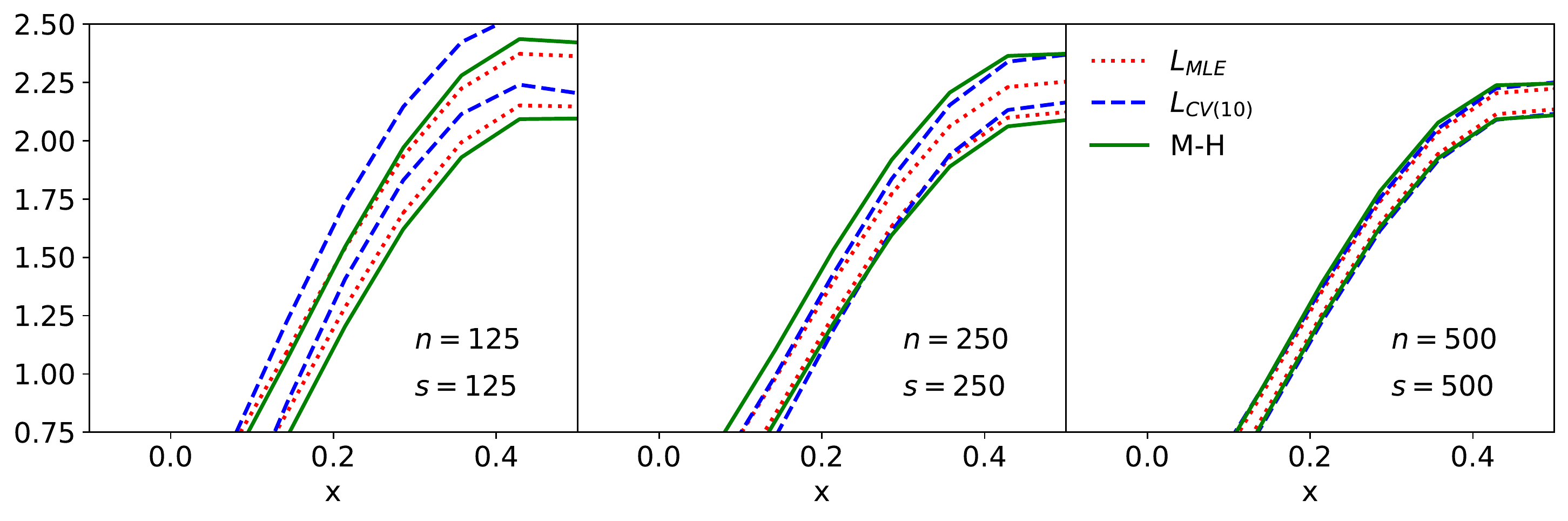}
		\caption{Details of $95 \%$ credible bands of posterior predictive distributions under the empirical Bayes approach and the fully Bayesian approach of Metropolis-Hastings algorithm. These were plotted at $t=1.00$.}
		\label{fig:Sim1:5:CB}
	\end{centering}
\end{figure*}

The following figures shows additional results of the empirical Bayes fit under the transverse harmonic wave simulation study at the time locations $t=0$, $t=0.43, t=0.71$, and $t=1$ (Figures \ref{fig:Sim1:2}, \ref{fig:Sim1:3}, \ref{fig:Sim1:4}, \ref{fig:Sim1:5}, \ref{fig:Sim1:2:CB}, \ref{fig:Sim1:3:CB}, \ref{fig:Sim1:4:CB}, and, \ref{fig:Sim1:5:CB}).

\end{appendices}

\end{document}